\documentclass[conference]{IEEEtran}
\IEEEoverridecommandlockouts
\usepackage{cite}
\usepackage{booktabs}
\usepackage[table,xcdraw]{xcolor}
\usepackage{amsmath,amssymb,amsfonts}
\usepackage{algorithm}
\usepackage{algpseudocode}
\usepackage{float}
\usepackage{textcomp}
\usepackage{url}
\usepackage{tabularx}
\usepackage{subcaption}
\usepackage{comment}
\usepackage{multirow}
\usepackage{etoolbox}
\usepackage{comment}
\usepackage{graphicx} 

\def\tablename{Table}
\usepackage{etoolbox}
\makeatletter
\patchcmd{\@makecaption}
  {\scshape}
  {}
  {}
  {}
\makeatletter
\patchcmd{\@makecaption}
  {\\}
  {.\ }
  {}
  {}
\makeatother
\def\tablename{Table}
\def\BibTeX{{\rm B\kern-.05em{\sc i\kern-.025em b}\kern-.08em
    T\kern-.1667em\lower.7ex\hbox{E}\kern-.125emX}}
\begin{document}

% The mapping of quantum algorithms in multi-core quantum computing architectures

\title{Adaptive Parallelism-Aware Qubit Routing for Ion Trap QCCD Architectures\\
}

\author{\IEEEauthorblockN{Anabel Ovide, Andreu Anglés-Castillo,  and Carmen G. Almudever}

\IEEEauthorblockA{\textit{Computer Engineering Department, Universitat Politècnica de València, Valencia, Spain}}}

\begin{comment}
\author{\IEEEauthorblockN{1\textsuperscript{st} Given Name Surname}
\IEEEauthorblockA{\textit{dept. name of organization (of Aff.)} \\
\textit{name of organization (of Aff.)}\\
City, Country \\
email address or ORCID}
\and
\IEEEauthorblockN{2\textsuperscript{nd} Given Name Surname}
\IEEEauthorblockA{\textit{dept. name of organization (of Aff.)} \\
\textit{name of organization (of Aff.)}\\
City, Country \\
email address or ORCID}
\and
\IEEEauthorblockN{3\textsuperscript{rd} Given Name Surname}
\IEEEauthorblockA{\textit{dept. name of organization (of Aff.)} \\
\textit{name of organization (of Aff.)}\\
City, Country \\
email address or ORCID}
\and
\IEEEauthorblockN{4\textsuperscript{th} Given Name Surname}
\IEEEauthorblockA{\textit{dept. name of organization (of Aff.)} \\
\textit{name of organization (of Aff.)}\\
City, Country \\
email address or ORCID}
\and
\IEEEauthorblockN{5\textsuperscript{th} Given Name Surname}
\IEEEauthorblockA{\textit{dept. name of organization (of Aff.)} \\
\textit{name of organization (of Aff.)}\\
City, Country \\
email address or ORCID}
\and
\IEEEauthorblockN{6\textsuperscript{th} Given Name Surname}
\IEEEauthorblockA{\textit{dept. name of organization (of Aff.)} \\
\textit{name of organization (of Aff.)}\\
City, Country \\
email address or ORCID}
}
\end{comment}

\maketitle

\begin{abstract}
Trapped-ion Quantum Charge-Coupled Device (QCCD) architectures promise scalability through interconnected trap zones and dynamic ion transport; however, this transport capability creates a complex compilation challenge: how to move qubits efficiently without degrading fidelity. We introduce a routing strategy that turns this challenge into an advantage by exploiting operational parallelism across traps while adapting to both algorithmic structure and device topology through a configurable multi-parameter scoring mechanism. Across a broad suite of benchmarks and QCCD layouts, the method consistently reduces ion-transport overhead and improves execution fidelity, outperforming state-of-the-art routing techniques. These results highlight that explicitly balancing movement overhead and execution parallelism under architectural constraints is key to unlocking the full potential of modular trapped-ion quantum processors.

\end{abstract}


\section{Introduction}

Quantum computing is redefining the limits of computation by exploiting superposition and entanglement, thereby enabling modes of information processing beyond the reach of classical systems. By using quantum state spaces that grow exponentially with the number of qubits, quantum processors promise advances in optimization, cryptography, and quantum simulation, among many other applications~\cite{8585034,Cao2019QuantumChemistry,365700,10.1145/237814.237866}. To physically realize these computational advantages, several hardware platforms have been developed, such as superconducting circuits, photonic qubits, neutral atoms, and trapped ion devices~\cite{Huang2020SuperconductingReview,RevModPhys.79.135,Henriet2020quantumcomputing,HAFFNER2008155}. Yet, as these systems scale beyond a few tens of qubits, maintaining performance while increasing architectural complexity has become one of the central challenges in the pursuit of practical, large-scale quantum computing.

%The scalability of quantum processors remains one of the most fundamental challenges in the realization of universal quantum computation. 
Among these quantum technologies, trapped-ion systems demonstrated remarkable levels of precision within single trapping zones~\cite{hughes2025trappediontwoqubitgates9999}. In particular, they stand out for their exceptional coherence times, high-fidelity gate operations, and precise qubit control~\cite{Wang2021,QMemory}. However, scaling a single-zone architecture becomes increasingly difficult, as the coupled motional modes and control interactions grow rapidly in complexity when many ions share the same potential~\cite{{PhysRevA.100.022332}}. To overcome these scalability constraints, the Quantum Charge-Coupled Device (QCCD)~\cite{Kielpinski2002} architecture was introduced as a modular approach for trapped-ion quantum computing. Rather than relying on a single monolithic trap, QCCD divides the system into multiple interconnected zones that allow ions to be physically moved between different traps via shuttling operations. This modularity preserves the high fidelities achievable in small traps while enabling the total qubit count to scale, establishing QCCD as one of the most promising pathways toward large-scale trapped-ion quantum processors.

The QCCD architecture increases complexity not only at the hardware level through the need to shuttle ions between zones, but also at the level of compilation. While single traps allow a straightforward compilation due to their all-to-all connectivity, ion trap modular systems require determining optimal ion movements across zones with the aim of minimizing the resulting transport overhead. At the same time, the QCCD architecture introduces new opportunities: its modular structure enables hardware-level parallelism by allowing independent operations to run simultaneously in separate trapping zones. Although parallel execution is, in principle, possible within a single trap, it is often limited by crosstalk that degrades gate fidelity~\cite{crosstalk}. QCCD mitigates these limitations, making parallelism a practical resource that can be exploited during compilation.

This capability enables ions to be distributed across zones to increase execution parallelism; however, doing so competes with the objective of minimizing shuttling and SWAP operations. Prior compilation approaches have predominantly focused on reducing ion movement to preserve execution fidelity~\cite{efficientcompilationshuttlingtrappedion,SYNC,compZonesMulti,muzzle}, yet they do not explicitly leverage the parallel-operation potential offered by modular architectures. However, as identified in~\cite{par}, balancing transport efficiency against parallel execution during routing can achieve up to 48\% improvement in execution fidelity for 40-qubit circuits. Therefore, explicitly accounting for the movement–parallelism trade-off is crucial when compiling circuits for QCCD-based ion-trap architectures.

In this work, we propose a new QCCD qubit routing algorithm that explicitly incorporates these considerations by combining ion-movement reduction with parallel-aware scheduling. The algorithm employs a heuristic search that not only identifies opportunities for operational parallelism but also adapts the routing process to both the structure of the quantum algorithm and the topology of the QCCD device through configurable weighting parameters that guide routing decisions. As a result, it achieves improved execution fidelity compared with state-of-the-art approaches, with an average gain of 56\% and enhancements of up to 120\% across evaluated benchmarks.

The remainder of this paper is organized as follows. Section~\ref{sec:back} introduces the background on single-trap and modular ion-trap architectures, as well as the compilation process for QCCD-based systems. Section~\ref{sec:art} summarizes current compilation methods from the state of the art and discusses operation parallelism in QCCD devices. Section~\ref{sec:parschedule} describes the proposed qubit routing algorithm. Section~\ref{sec:met} presents the benchmark selection, topology configurations, evaluation metrics, and execution framework. Section~\ref{sec:res} evaluates the performance of the routing strategy through a series of experiments. Finally, Section~\ref{sec:con} concludes the paper and outlines future research directions.
\section{Background}
\label{sec:back}
The implementation of scalable quantum computers requires reliable qubit control, high-fidelity operations, and architectures capable of supporting large-scale connectivity and error correction. Trapped-ion systems satisfy these requirements, offering long coherence times and precise qubit control. This section reviews single-zone ion-trap architectures, introduces the Quantum Charge-Coupled Device as a scalable extension, and outlines the compilation challenges of modular ion-trap systems.

\subsection{Monolithic architecture: single ion traps}
In a typical trapped-ion quantum processor, ions are confined in a linear or surface-electrode Paul trap and manipulated using coherent electromagnetic fields~\cite{Wineland1998}. Quantum operations exploit their internal electronic states and shared vibrational motion to mediate interactions between qubits. Two principal schemes are commonly used for such operations: the \textit{Cirac–Zoller gate}~\cite{Cirac1995}, and the \textit{Mølmer–Sørensen gate}~\cite{Molmer1999,HAFFNER2008155}. State-of-the-art implementations of both single and two-qubit gates have achieved fidelities of 99.99\%~\cite{hughes2025trappediontwoqubitgates9999}. Although parallel gate execution is theoretically possible within a single trap, simultaneous multi-qubit operation introduces additional motional-mode coupling that can lead to measurable crosstalk-induced fidelity degradation~\cite{crosstalk}.

Within a single trapping zone, ions provide native all-to-all connectivity, allowing any pair of qubits to interact directly and eliminating the routing constraints present in fixed-connectivity architectures such as superconducting circuits~\cite{Huang2020}. Trapped-ion qubits are intrinsically identical~\cite{Ballance2016}, exhibit long coherence times, making them suitable as reliable quantum memories and communication nodes~\cite{Wang2021,QMemory}, and offer high-fidelity state readout~\cite{ITMeasurement}. In addition, their compatibility with photonic interfaces enables distributed quantum computing, with remote entanglement demonstrated over distances exceeding 200 meters~\cite{ITNetwork,200entanglement}. 

IonQ’s upcoming Tempo system reports an algorithmic-qubit benchmark of \#AQ 64 and target two-qubit gate fidelities approaching 99.9\% ~\cite{IonQTempo2025}. IonQ has also demonstrated quantum error-correction protocols in long trapped-ion chains, supporting logical encoding and fault-tolerant operation within single-trap architectures~\cite{Ye2025,ecit,Postler2022}. Similarly, academic groups at the University of Maryland and the University of Innsbruck have demonstrated high-fidelity multi-qubit control and experimental fault-tolerant operations~\cite{UMD2023,Innsbruck2024}.

While single-zone ion traps provide excellent performance for small qubit systems, scaling them introduces fundamental physical constraints. As more ions are added to a single trap, their shared motion becomes increasingly complex, and motional heating increases, making precise control and high-fidelity operations harder to maintain~\cite{PhysRevA.100.022332}. Since this heating directly impacts system fidelity~\cite{paulNoise}, single-zone architectures face practical limits in size, motivating the development of modular and more scalable multi-zone trap designs such as the QCCD model.

\subsection{Modularity for scaling ion-trapped-based hardware}

The QCCD architecture was introduced as a modular extension of conventional ion traps to overcome the scalability and control limitations inherent to single-zone systems~\cite{Kielpinski2002, raceTRack}. In this design, trapping electrodes are segmented into independently controlled regions, forming a multi-trap architecture capable of confining, storing, and manipulating ions in separate zones. Ions can be dynamically transported between these regions by adjusting the voltages applied to the electrodes. To accomplish this, the process involves three steps: (i) \textit{splitting}, which spatially disentangles a single ion from an ion crystal; (ii) \textit{shuttling}, which moves the ion from one zone to another; and (iii) \textit{merging}, which integrates the ion back into an ion crystal.

Several experimental platforms have realized functional QCCD systems. For instance, Quantinuum’s H-Series processors implement multi-zone layouts enabling high-fidelity gate operations and ion shuttling across (intersections between trap segments that enable ion shuttling). Their most recent device, Helios, achieves 98 qubits with single-qubit fidelities of 99.99\% and two-qubit fidelities of 99.21\%~\cite{QuantinuumHelios2025}, with further improvements expected~\cite{QuantinuumFT2025}. The University of Innsbruck and AQT have demonstrated two-dimensional architectures such as the Quantum Spring Array (QSA), enabling scalable inter-site interactions without merging ion chains~\cite{b9s1-6r44}.

Beyond QCCD, additional strategies, including higher-dimensional ion crystals~\cite{Bruzewicz2016} and photonic interconnects linking multiple QCCD chips~\cite{PhysRevX.15.011040}, have been explored to increase qubit capacity. Recent work on ion–photon interfaces~\cite{Akhtar2023}, compiler systems such as TITAN~\cite{titan}, and integrated control-electronics platforms targeting thousand-qubit devices~\cite{1kqubitIT} further support large-scale integration.

As these architectures scale, fault tolerance becomes increasingly important. Recent quantum error correction demonstrations~\cite{paetznick2024demonstrationlogicalqubitsrepeated, tham2025distributedfaulttolerantquantummemories, yin2025flexionadaptiveinsituencoding} further highlight that scalable ion transport and reliable zone coordination are essential for future trapped-ion architectures.

Nevertheless, the flexibility of modular architectures introduces new challenges for its compilation process. Efficient execution of quantum circuits requires optimizing the sequence of ion movements, gate placements, and zone utilization to minimize transport overhead and decoherence as shuttling operations induce motional heating that directly reduces gate fidelity~\cite{paulNoise}. Consequently, the design of routing and compilation algorithms specifically tailored to QCCD systems has become a crucial research area, directly influencing circuit runtime, parallelization capability, and overall fidelity.

\subsection{Compilation for Modular QCCD Architectures}
The compilation of quantum algorithms translates hardware-agnostic circuits into executable programs~\cite{comp_surv}. The compilation workflow generally comprises four main stages: (i) \textit{gate decomposition}, which decomposes gate operations into the native gate set supported by the hardware; (ii) \textit{qubit mapping}, where logical qubits in the circuit are assigned to physical qubits; (iii) s\textit{cheduling}, which parallelized quantum gates for the execution; and (iv) \textit{qubit routing}, which introduces the necessary operations or movements to ensure that non-adjacent qubits can interact. Each quantum technology imposes its own architectural and physical constraints, and therefore requires specialized compilation methods tailored to its particular capabilities and limitations~\cite{NN,10.1145/3655029,zhu2025quantumcompilerdesignqubit}.

\begin{figure}[t!]
  \centering
  \hspace*{-0.5cm}  % adjust this value
  \includegraphics[width=1.1\columnwidth]{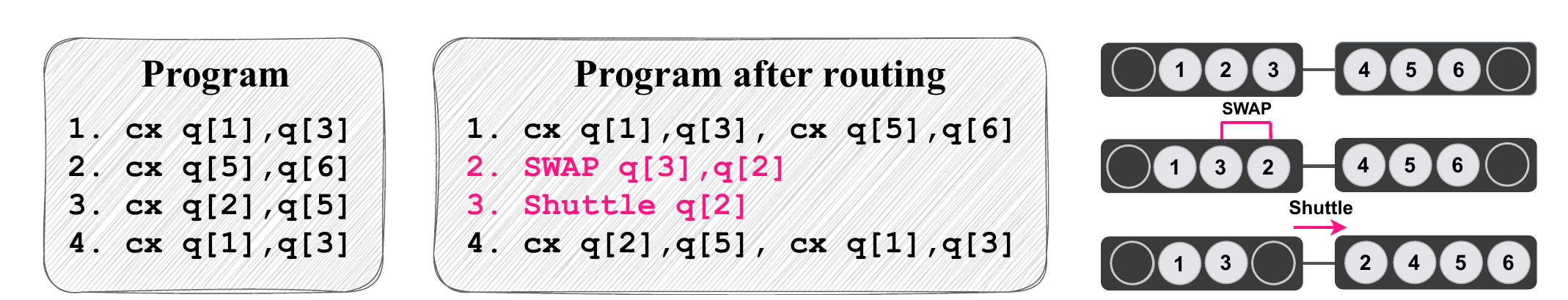}

    \caption{\small{Example of compiling a quantum circuit onto a 1D linear topology consisting of two 4-ion capacity traps. Left: original circuit. Center: compiled circuit with an added SWAP and Shuttle. Right: compiling process (from top to bottom), (i) initial allocation of three ions in each trap, enabling the first two CX gates (executed in parallel), (ii) SWAP to move qubit 2 to the trap edge, and (iii) Shuttle to the adjacent trap to perform the remaining CX gates.}}
    \label{fig:Mappingexample}
\end{figure}

In conventional linear ion traps, qubits exhibit all-to-all connectivity within a single trapping zone. This eliminates the need for complex qubit mapping or routing procedures, allowing the compilation workflow to focus primarily on gate transpilation and operation scheduling. As a result, the overall compilation process in single-zone ion traps is comparatively straightforward.

However, in QCCD-based ion-trap architectures, compilation becomes significantly more challenging. Since ions are distributed across multiple trapping zones, interactions between them often require physically relocating ions, typically through shuttling and SWAP operations, which introduce additional errors and reduce execution fidelity. Therefore, minimizing ion movement operations becomes a key objective of the compiler.

To address this, logical qubits must be mapped efficiently from the virtual circuit to the physical device~\cite{im,10.1145/3526241.3530366}, as a good initial placement can significantly reduce the number of required ion movements. In addition, qubit routing becomes critical:ions must be transported across the device so that interacting qubits can be co-located while respecting topology constraints such as trap congestion and transport paths in the QCCD architecture~\cite{efficientcompilationshuttlingtrappedion, SYNC,cycles,compZones, compZonesMulti,muzzle}. An example of a qubit routing procedure is shown in Figure~\ref{fig:Mappingexample}.

Moreover, operation scheduling becomes more complex in QCCD systems. Unlike single-zone traps, where usually only one gate can be executed at a time due to crosstalk errors, multi-zone architectures allow for concurrent operations in different trapping regions. This increased parallelism requires a more sophisticated scheduling strategy to exploit available resources while avoiding conflicts in ion movements and zone occupancy.

\section{Current Landscape of QCCD Compilation}
\label{sec:art}
Efficient compilation is essential for fully exploiting the capabilities of modular trapped-ion architectures such as QCCD systems. Although prior work has introduced a variety of routing and scheduling strategies, these approaches largely concentrate on transport optimization and conflict mitigation, leaving the inherent parallelism of modular architectures underutilized. Because concurrent operations across independent zones can substantially increase fidelity and reduce both execution time and circuit depth, effectively utilizing this parallelism becomes a central consideration for scalable compilation. The following subsections review existing quantum circuit compilation strategies and examine the impact of parallel execution within QCCD qubit routing.

\subsection{State of the art}

Recent work on modular trapped-ion architectures has explored various compilation strategies for multi-trap and distributed QCCD systems. The approach in~\cite{multiQCCD} introduces the entanglement-module-linked QCCD (EML-QCCD), where multiple QCCD units are interconnected via optical fibers, enabling photonic remote entanglement. Their compiler performs routing both within each QCCD and between modules. To mitigate multi-module congestion, the method employs a qubit-replacement policy analogous to classical LRU (Least Recently Used), shuttling out the qubit that has remained idle the longest . Zone selection is also performed heuristically, favoring available regions closest in hierarchical level, and the same prioritisation mechanism is used to resolve conflicts when multiple qubits compete for placement.

A different direction is taken by Bach et al.~\cite{efficientcompilationshuttlingtrappedion}, which proposes the Position Graph abstraction to unify trapped-ion hardware with graph-based mapping strategies commonly used for superconducting devices. In this representation, nodes encode ion positions and edges correspond to physical transitions such as move, split, and merge. Building on this abstraction, their SHAPER algorithm uses a heuristic cost function to minimise shuttling distance and congestion.

The S-SYNC framework~\cite{SYNC} introduces a generic-swap primitive encompassing all QCCD actions (move, split, merge, and SWAP). By formulating a weighted connectivity graph that encodes movement, congestion, and gate-ordering penalties, S-SYNC performs heuristic search guided by penalties on movement, repeated swaps, and local congestion.

A complementary class of approaches examines cycle-based shuttling. Schoenberger et al.~\cite{cycles} treat the QCCD as a graph and exploit closed loops to shuttle multiple ions in parallel. This idea is extended in~\cite{compZones} to systems with separate memory and processing zones, and later generalised to multi-processing-zone layouts in~\cite{compZonesMulti}. 

Finally, the work of Ash Saki et al.~\cite{muzzle} introduces three targeted optimisations: a future-operations-based shuttle-direction policy to avoid repeated back-and-forth shuttling, opportunistic gate reordering to proactively free capacity in destination traps, and a nearest-neighbour-first rebalancing routine to resolve shuttle-path traffic blocks efficiently.

Although these approaches advance compilation for trapped-ion systems, they largely focus on avoiding congestion or reducing transport cost, and none explicitly aim to maximize the operational parallelism enabled by the QCCD architecture.

\subsection{Operation parallelization trade-off}

As previously discussed, modular ion trap architectures enable parallel execution across multiple trapping zones, reducing circuit depth through simultaneous gate operations. However, taking advantage of this parallelism can require additional shuttling and SWAP operations, adding transport overhead that may offset potential benefits. Consequently, effective use of parallelism requires balancing depth reduction with the ion-movement costs introduced.

The work in~\cite{par} investigates the trade-off between operation parallelization and ion-movement overhead in QCCD architectures. The results show that, as devices scale, a moderate degree of parallelization can significantly improve final execution fidelity by reducing circuit depth, with improvements of up to 48\% for 40-qubit circuits. However, excessive parallelization has the opposite effect: aggressively distributing ions across multiple traps leads to increased shuttling and SWAP operations, which ultimately degrade fidelity. Thus, an optimal balance must be found between executing operations sequentially and maximizing parallel execution opportunities.

Moreover, the benefits of parallelization depend strongly on additional factors such as the structure of the quantum algorithm and the specific topology of the QCCD device. Different circuits exhibit different qubit interaction patterns, and device layouts vary in their connectivity and transport characteristics. These observations indicate that qubit routing and operation scheduling strategies must be flexible and tunable to the particular algorithm–architecture combination in order to consistently achieve high-fidelity execution.

Despite this, state-of-the-art qubit routing approaches generally do not account for the parallelization capabilities of modular architectures when planning ion movements. To address this limitation, we develop a routing strategy that explicitly incorporates parallel-execution opportunities, along with the associated transport and architectural constraints. This approach can adapt to different quantum algorithms as well as to various QCCD ion-trap topologies, enabling more efficient executions and improved overall fidelity.

\section{Adaptive Parallelism-Aware Qubit Routing and Scheduling}

\label{sec:parschedule}

\begin{figure*}[!t]
  \centering
  \includegraphics[width=1.05\textwidth]{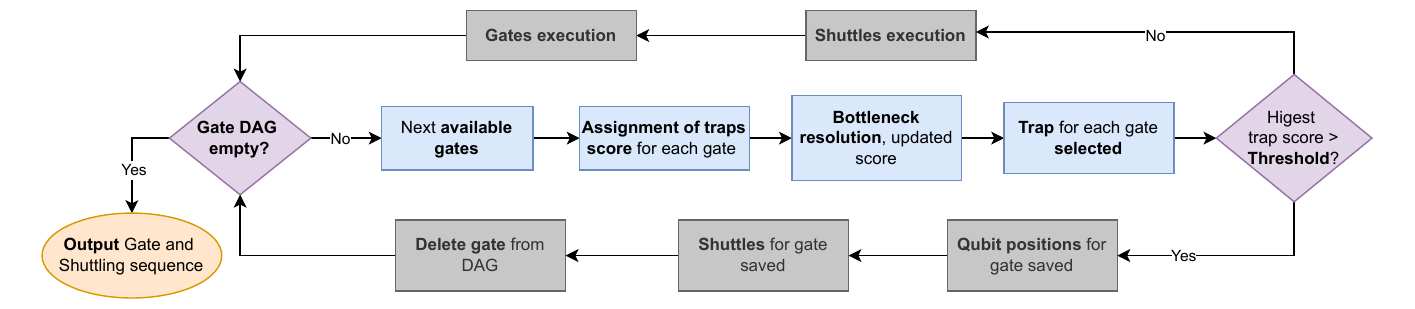}
  \caption{\small{Flow diagram illustrating the steps of the proposed routing algorithm.}}
  \label{fig:diagram}
\end{figure*}

In this section, we introduce a qubit routing and gate scheduling method designed to exploit the parallelism execution of the different operations while adapting to the specific requirements of different quantum algorithms and QCCD topologies. The approach incorporates several key factors: (i) minimizing qubit movement by reducing shuttling and SWAP operations, (ii) accounting for upcoming operations to improve routing decisions, (iii) preventing bottlenecks by managing congestion across traps, and (iv) maximizing parallel execution whenever possible.

\subsection{Routing Algorithm Overview}

Our approach is based on an adaptive, parallelism-aware routing and scheduling strategy tailored to QCCD-based modular ion-trap architectures. It makes use of graph-based representations of both the hardware topology and circuit dependencies to guide coordinated routing decisions. By jointly considering ion-transport cost, trap capacity constraints, and gate-level parallelism, the method dynamically balances movement overhead and execution concurrency to improve overall circuit fidelity and depth.

The QCCD architecture is represented as a graph in which traps serve as nodes that store ions, and junctions form edges representing allowed ion-transport paths. This abstraction enables efficient computation of key primitives such as shortest-path queries, allowing the routing algorithm to operate on arbitrary device topologies.

Quantum circuits are encoded using a gate-level Directed Acyclic Graph (DAG), where dependencies arise from shared qubits; gates with no predecessors form the set of currently executable operations. A separate shuttling DAG tracks transport operations, with dependencies created when multiple ion shuttling require the same junction or otherwise conflict. This representation enables parallel transport whenever allowed by the device topology.

The overall workflow, illustrated in Fig.~\ref{fig:diagram}, begins by identifying all executable gates in the gate DAG. For each gate, all traps on the shortest path between the participating ions are evaluated and assigned a score, and the highest-scoring trap is selected as the candidate execution site. Any bottlenecks detected during scoring are resolved immediately, and trap scores are updated accordingly. After all available gates have been scored, the gate with the highest trap score is selected. If its score is below a predefined threshold, the gate is deferred and pending operations from earlier rounds are executed instead. Once selected, the gate’s ions are assigned to the chosen trap, the required shuttling operations are generated and recorded, and gates sharing the same qubits are temporarily removed and reconsidered in later rounds. The updated ion positions are stored in the machine state.

Once no gates remain in the DAG without unresolved dependencies, or the highest gate score falls below the threshold, all stored shuttling operations (including SWAPs) are executed in parallel over one or more rounds. After the ions reach their updated positions, all gates assigned to the same time step are applied concurrently within a single scheduling cycle. This procedure repeats until the gate DAG is empty.

\begin{comment}
\begin{algorithm}[H]
\caption{Generic QCCD Routing Algorithm}
\label{alg:qccd_routing}
\begin{algorithmic}[1]
\State \textbf{Input:} Quantum circuit gates $G$, trap topology $T$
\State \textbf{Output:}  Sequence of trap assignments  $\mathcal{T}$ and shuttling operations  $\mathcal{M}$

\ForAll{$g_i=(q_1,q_2) \in G$}
    \State Compute shortest paths $\mathcal{P}_{\min}(T(q_1),T(q_2))$
    \Statex 
    \Comment{$S(T_j)$: trap score function}
    \State Evaluate $S(T_j)$ for all $T_j \in \mathcal{P}_{\min}(q_1,q_2)$
    \State Solve possible bottlenecks and update score
    \State $T^* \gets \operatorname*{argmax}_{T_j} S(T_j)$ \Comment{Trap with highest score}

    \If{$S(T^*) < \tau$} \Comment{$\tau$: score threshold}
        \State Discard $g_i$ for next round
        \State Execute shuttles in $\mathcal{M}$
        \State Execute previously scheduled gates in  $\mathcal{T}$
    \Else
        \State Schedule $g_i$ on trap $T^* \in  \mathcal{T}$
        \State Assign qubits $\{q_1,q_2\}$ to $T^* \in  \mathcal{T}$
        \State Record shuttle operations in $\mathcal{M}$
        \State Prune from $G$ gates acting on qubits of $g_i$
        \State Update machine configuration $\mathcal{C}$
    \EndIf
\EndFor
\Statex
\end{algorithmic}
\end{algorithm}
\end{comment}

\subsection{Routing Cost Function}

To determine the most appropriate trap for executing a gate, a score is computed for each candidate trap based on five parameters. The first two, \textit{shuttles} (SH) and \textit{SWAPs} (SW), contribute negative values, as the objetive is to minimize ion movement. SH denotes the number of shuttling operations required for the ion to reach the destination trap, while SW indicates the number of SWAPs needed to bring the ion to the end of a trap so that it can be shuttled. These two parameters are strongly correlated; in most situations, each required shuttle (unless the ion is already at the trap edge) necessitates one or more SWAPs. 

The next parameter, \textit{future operations} (FO), considers upcoming gates involving the same qubits in a trap. If a qubit scheduled for a future interaction shares a trap with its partner qubit, a positive value is added to the score. Earlier future operations contribute more heavily than later ones, thereby reducing unnecessary ion movement by co-locating qubits in advance. The score for FO for a given trap is computed as follows:

\begin{equation}
\hspace*{-0.15cm}FO(q, T) =\;
\sum_{i=1}^{L} \big(L - i\big)
\sum_{g \in G_i}
\mathbf{1}\!\Big[
(q \in Q(g)) \,\wedge\,
(Q(g) \cap T \neq \emptyset)
\Big].
\label{eq:future_op_score}
\end{equation}

\noindent
In this expression, $FO(q, T)$ denotes the future-operation score for qubit $q$ with respect to trap $T$. The parameter $L$ represents the number of circuit layers considered in the look-ahead window (set to 7 in our experiments). Each layer $l$ contains a set of gates $G_l$, and for any gate $g$, the set $Q(g)$ identifies the qubits on which the gate acts. The indicator function $\mathbf{1}[\cdot]$ evaluates to 1 when its condition is satisfied and 0 otherwise.

Eq.~\ref{eq:future_op_score} accumulates the contributions of all future gates that involve qubit $q$
and at least one ion from trap $T$. Each contribution is weighted by $(L - i)$, meaning that gates located deeper in time
(i.e., closer to the end of the time window) have reduced influence on the total score, while earlier gates contribute more strongly.

Another important parameter is the \textit{Excess Capacity} (EC), which measures the number of free spots in a trap. A negative score is assigned when the trap is full, since accommodating an additional qubit would require reallocating ions and therefore introducing extra movements. Conversely, a positive contribution is added when the trap has available capacity, with larger values assigned to traps offering more free space. The final parameter, \textit{Parallelism} (PR), provides a positive contribution when the trap has no gate scheduled for the current time slice, encouraging the algorithm to exploit parallel execution, and a negative contribution otherwise.

In summary, the score assigned to each trap is determined by five parameters: \textit{Shuttles} (SH), \textit{SWAPs} (SW), \textit{Future Operations} (FO), \textit{Excess Capacity} (EC), and \textit{Parallelism} (PR). It is important to note that these contributions assign equal weight to all parameters, although in practice certain algorithms or topologies may benefit from weighting specific terms more strongly than others.

\begin{algorithm}[h!]
\caption{Trap Selection for a Two-Qubit Gate}
\label{alg:trap_selection}
\begin{algorithmic}[1]
\State \textbf{Input:} Pair of qubits $(q_1,q_2)$ to execute a 2q gate $\mathcal{G}$
\State \textbf{Output:}Selected trap $T^*$, updated ion configuration $\mathcal{C}$, and ion movements $\mathcal{M}$

\If{$T(q_1)=T(q_2)$}
    \State Compute trap score $S(T(q_1))$ using Eq.~\ref{ec:TrapScore}
\Else
    \State Compute shortest paths $\mathcal{P}_{\min}(T(q_1),T(q_2))$

    \ForAll{$P=(T_1,\ldots,T_r) \in \mathcal{P}_{\min}(T(q_1),T(q_2))$}
        \ForAll{$T_j \in P$} \Comment{\textit{Shuttle} and \textit{SWAP} costs}
            \If{$|\{q_1,q_2\}\cap T_j| = 1$}
                \State Number of ion movements for $\{q_1,q_2\}\setminus T_j$

            \ElsIf{$|\{q_1,q_2\}\cap T_j| = 0$}
                \State Number of ion movements for both ions
            \EndIf

            \If{$T_j \in \mathcal{B}$} \Comment{$\mathcal{B}$: bottleneck traps}
                \State Resolve via Alg.~\ref{alg:bottleneck_resolution}
            \EndIf
            \State Calculate Trap score $\mathcal{S}_{T_j}$ by Eq.~\ref{ec:TrapScore}
        \EndFor
    \EndFor
    \State Final trap $T^*=\operatorname*{argmax}_{T_j\in P,\;P\in\mathcal{P}_{\min}(q_1,q_2)} s(T,P)$
\EndIf
\Statex
\If{$\mathcal{S}(T^*) \geq \tau$} \Comment{Threshold comparison}
    \State Save gate to execute  $\mathcal{G}$
    \State Update ion configuration $\mathcal{C}$
    \State Record shuttle operations $\mathcal{M}$
\Else
    \State Postpone gate to next time slice
\EndIf

\State \Return $T^*$, $\mathcal{C}$, $\mathcal{M}$

\end{algorithmic}
\end{algorithm}

For instance, when executing a quantum algorithm with many two-qubit gates per time slice, it is often preferable to prioritize minimizing shuttles and SWAPs, since ion movement within the traps is already significant. In such cases, assigning higher weight to SH and SW helps avoid additional transport, whereas prioritizing parameters such as parallelism could distribute qubits across traps and further increase ion movement. Conversely, for algorithms with relatively few two-qubit gates per time slice, increasing the weight of parallelism may be more beneficial, as it enables greater concurrency without significantly raising the movement overhead. Similar considerations apply to the topology of the hardware: if the architecture contains complex or costly junctions, it may be preferable to prioritize minimizing shuttling over other parameters. To accommodate these variations, each scoring parameter is assigned a configurable weight that can be adapted to both the quantum algorithm and the device topology. The final score is computed as:

\begin{equation}
\label{ec:TrapScore}
    \textit{S(q,T)} = \alpha\text{SH} +  \lambda\text{SW}  +  \beta\text{FO}   +  \sigma\text{EC} +  \gamma\text{PR}.
\end{equation}

Eq.~\ref{ec:TrapScore} is used when assigning a pair of qubits to a trap for gate execution. However, when the score must be computed for an ion that needs to be relocated to another trap due to a bottleneck (i.e., a congestion condition in which the target trap has reached its ion-capacity limit), the scoring procedure is simplified:

\begin{equation}
\label{ec:TrapScoreB}
    \textit{S}_{\text{B}}\textit{(q,T)} = \text{SH} +  \text{SW}  + \text{FO}.
\end{equation}

For this computation, parameter weights are not applied and the \textit{Parallelism} component is omitted, since the ion is moved solely for resolving a bottleneck rather than for executing a gate. Similarly, the \textit{Excess Capacity} term is excluded, because the target trap has been verified to provide sufficient capacity, as elaborated in later sections.

\subsection{Trap selection}

\begin{figure}[t!]
  \centering

  \includegraphics[width=1.08\columnwidth]{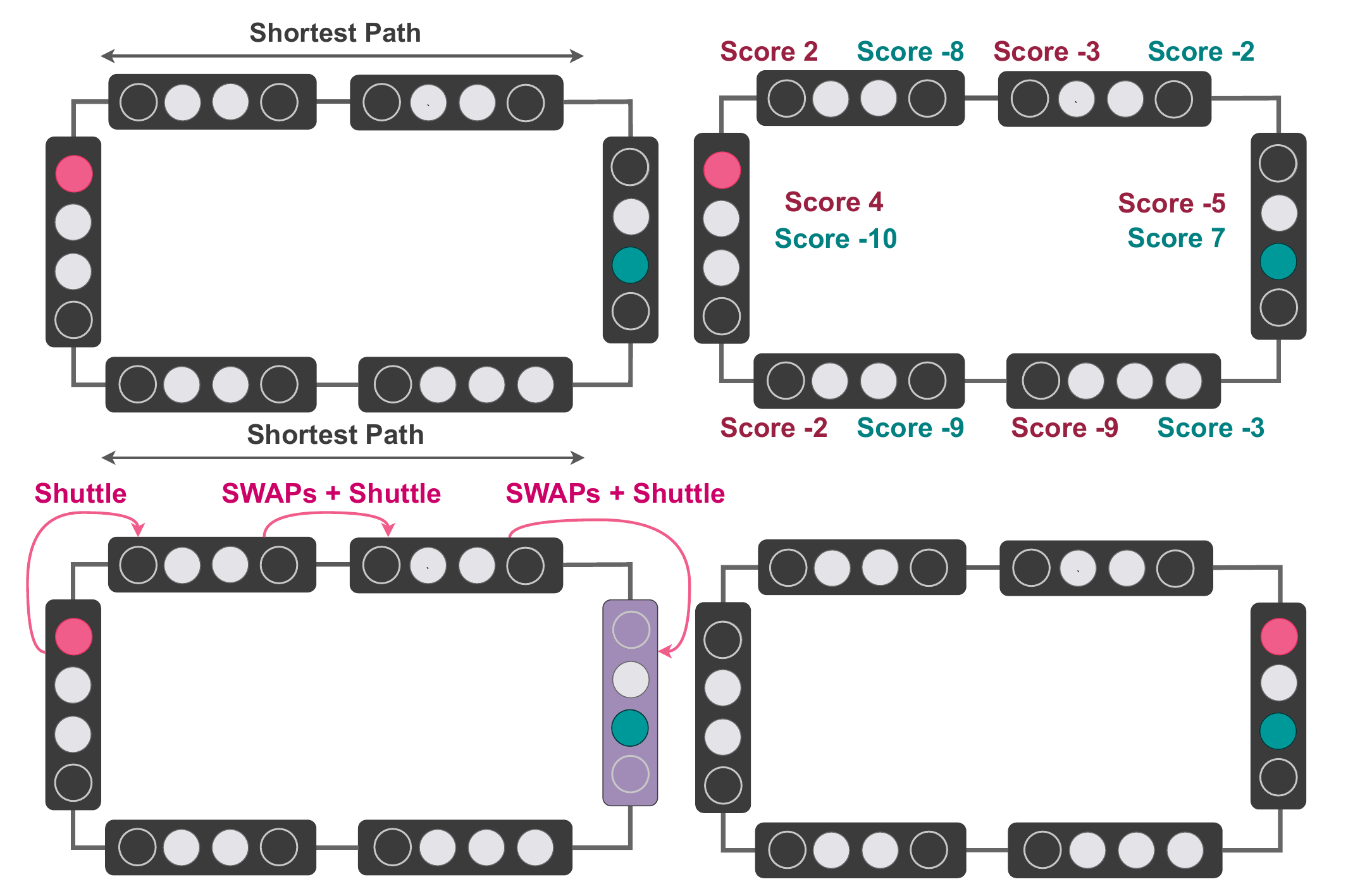}

  \caption{\small{Example of the trap selection process. (i) Shortest paths are computed (top left), (ii) trap scores are evaluated (top right), (iii) the highest-scoring trap is selected and ion movements applied (bottom left), (iv)  and the resulting movements and positions are recorded (bottom right).}}
  \label{fig:trapex}
\end{figure}

To determine the trap in which a pair of qubits will execute a two-qubit gate, the algorithm first checks whether both qubits are already located in the same trap. If so, the score for that trap is computed using Eq.~\ref{ec:TrapScore} and recorded (Algorithm~\ref{alg:trap_selection}, lines 3-4).

If the qubits are not co-located, all shortest paths between their current traps are computed (Algorithm~\ref{alg:trap_selection}, line 6). When multiple shortest paths exist, each path is evaluated independently (line 7). This is illustrated in the example shown at the top-left of Fig.~\ref{fig:trapex}, where the pink and blue ions require co-location to execute their gate, and two distinct shortest paths allow this movement. For every trap along these paths, including the traps currently holding the qubits, the score from Eq.~\ref{ec:TrapScore} is calculated  (line 9). The trap with the highest score is then selected as the execution location for the gate.

The shuttle and SWAP scores for each candidate trap are obtained by counting the ion movements required to bring the qubits to that trap. If one qubit is already present, only the movements of the other are considered (line 10); otherwise, the movements of both are included (line 11). SWAPs accumulate as ions traverse intermediate traps, where each trap contributes a number of SWAPs equal to the number of ions it currently holds. If a bottleneck is encountered and additional ions must be relocated, the score is updated accordingly, as described in the next subsection  (lines 14-15). The top-right panel of Fig.~\ref{fig:trapex} illustrates this effect: fewer ions per trap along the upper path reduce the SWAP cost and increase the Excess Capacity contribution, resulting in higher overall scores.

Once all traps along the shortest paths have been evaluated, the trap with the highest score is selected (line 20), and the corresponding shuttling route and final ion positions are recorded. In the example, the trap initially containing the blue ion obtains the highest score and is therefore selected as the execution trap. The pink ion is then routed along the chosen shortest path to that trap.

The selected trap score is then compared against a threshold, a tunable hyperparameter that determines the minimum acceptable trap score for committing a gate execution in the current time slice (line 22). This threshold controls the balance between parallel and sequential execution. A low threshold allows traps with larger movement costs (i.e., more negative scores) to be accepted, increasing parallelism at the expense of additional ion movement. In contrast, a high threshold restricts execution to traps requiring minimal movement, yielding a more sequential behavior. If the selected trap score satisfies the threshold, the machine state is, and all required shuttling operations are stored (lines 23-25).

\subsection{Bottleneck Resolution}

\begin{figure}[t!]
  \centering

  \includegraphics[width=0.8\columnwidth]{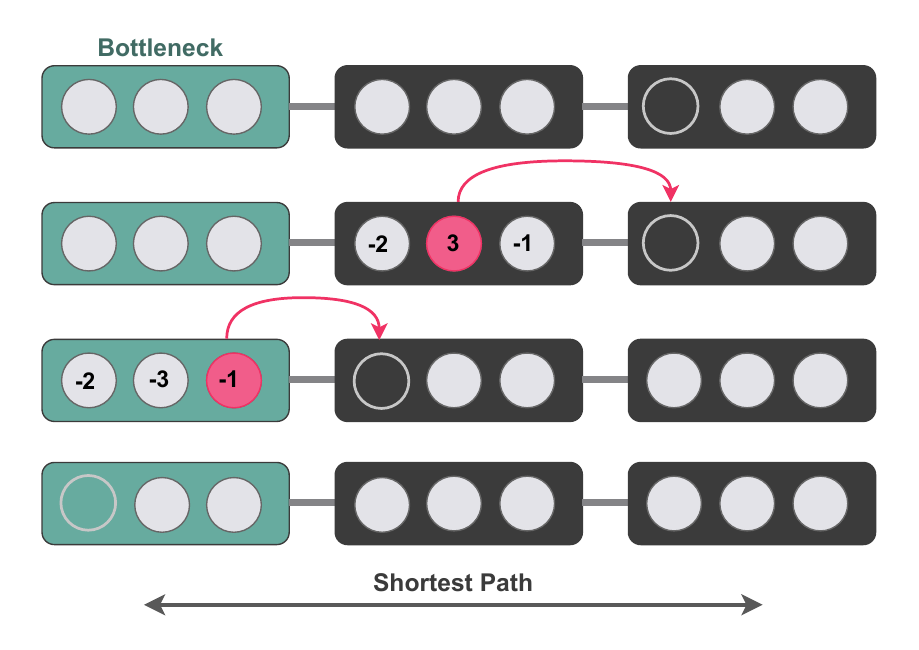}

  \caption{\small{Example of congested-trap resolution (top to bottom). (i) The trap highlighted in green must free one slot ($EC=0$), but the adjacent trap along the relocation path is also full. The shortest path to the nearest trap with available capacity ($EC>0$) is identified. (ii) Ions in the adjacent trap are scored using the relocation metric, and the highest-scoring movable ion is selected. (iii) The selected ion is shuttled to the trap with free capacity, propagating space backward along the path. (iv) All ion movements and updated positions are recorded.}}
  \label{fig:bottle}
\end{figure}

When an ion must be moved to a different trap, it may need to traverse a sequence of intermediate traps. In some situations, one or more of these traps may be full, meaning no additional ion can be allocated. In such cases, at least one ion from the congested trap must be relocated to free a space. This process is delicate, as it can significantly increase the total number of ion movements and may force ions into non-optimal traps, potentially degrading routing performance.

To optimally address this situation, when a bottleneck is detected, the algorithm first checks whether any ion in the congested trap can be moved out. In some cases, all ions in that trap are scheduled for imminent operations and therefore cannot be relocated. If none of the ions can be moved, the routing path that passes through the bottleneck is rejected; since multiple shortest paths may exist, the algorithm evaluates all remaining alternatives. 

If at least one ion can be moved, the algorithm computes all shortest paths to the nearest trap with available capacity. For each path, the algorithm evaluates and stores the relocation scores associated with resolving the bottleneck, ensuring that all possible resolution strategies are considered and selects the feasible path that yields the highest score (Algorithm~\ref{alg:bottleneck_resolution}, line 3).

This situation is illustrated in Fig.~\ref{fig:bottle}, where the leftmost trap is full and must release one of its ions. To resolve this, the algorithm identifies the nearest trap with available capacity (the rightmost trap in the figure). If any of the intermediate traps between the bottleneck and the destination trap also forms a bottleneck and none of its ions can be moved because they are all involved in pending operations, the corresponding shortest path is discarded, and the algorithm evaluates the remaining candidate paths.

\begin{algorithm}[h!]
\caption{Bottleneck Resolution}
\label{alg:bottleneck_resolution}
\begin{algorithmic}[1]
\State \textbf{Input:}  Bottleneck trap $T_j$, ion configuration $\mathcal{C}$
\State \textbf{Output:} $\mathcal{C}$, ion movements $\mathcal{M}$ and score $\mathcal{S}_{best}$

\State Shortest paths $\mathcal{P}_{\min}(T_j) = \{\,P=(T_1,\dots,T_j)\}$  to the nearest traps with $\operatorname{EC}> 0$

\ForAll{$P \in \mathcal{P}_{\min}(T_j)$}
    \State Reverse path $P^* = (T_{m_r}, T_{m_{r-1}}, \dots, T_j)$
    \For{$T_{m_i} \in P^* \setminus \{T_{m_r}\}  $}
                \If{$\operatorname{EC}(T_{m_i}) = 0$}\Comment{No free spaces in trap}
                                    \State $\mathcal{Q}_m = \{\, q_x \in T_m : q_x \text{ is movable}\,\}.$

                            \If{$\mathcal{Q}_m = \emptyset$}
                \State \textbf{continue} to next path $P'$
            \EndIf
            \State For each $q_x \in \mathcal{Q}_m$, compute score Eq.~\ref{ec:TrapScoreB}.
            \State $q^* \gets \operatorname*{argmax}_{q_x} s(q_x,T_{m+1})$
            \State Move $q^* \rightarrow T_{m-1}$
            \State Record score ions and movements in $\mathcal{S}_P$
            \If{$\mathcal{S}_{{best}} < \mathcal{S}_P$}
                \State Update ion configuration $\mathcal{C}_P$
                \State Record movement in $\mathcal{M}_P$
                \State $\mathcal{S}_{{best}} \gets \mathcal{S}_P$
            \EndIf

        \EndIf
    \EndFor
\EndFor

\If{$\mathcal{S}_{{best}} = \emptyset$} \Comment{No available paths}
    \State Postpone gate to next time slice; \Return
\EndIf

\State \Return $\mathcal{C}, \mathcal{M}, \mathcal{S}_{{best}}$
\end{algorithmic}
\end{algorithm}

To decide which ion should be moved into the adjacent trap, the algorithm first identifies which ions in the bottleneck trap are available to be moved. For each movable ion, the bottleneck score defined in Eq~\ref{ec:TrapScoreB} is computed relative to the adjacent trap, and the ion with the highest score is selected for movement (Algorithm~\ref{alg:bottleneck_resolution}, lines 12-13).

During this backward propagation of free space, each intermediate trap along the path must also be able to release at least one ion. If one intermediate trap cannot relocate an ion because all ions are involved in scheduled operations, that path is discarded, and the next shortest path is evaluated. If no remaining path satisfies this condition, the gate is postponed to the next time slice (lines 9-11)

Continuing with the example of Fig.~\ref{fig:bottle}, the trap closest to the one containing free capacity evaluates its ions, selects the highest-scoring candidate, and relocates it into the available slot. The same procedure is then applied to the bottleneck trap, ultimately freeing the necessary space.

Once a movable ion has been selected, the algorithm records its score, the required shuttling steps, and the updated ion position in the machine graph (line 15). After the bottleneck is resolved, the accumulated score for that path is compared against the scores obtained for other shortest-path resolutions (if any). If the current resolution produces a higher total score, it is retained; otherwise, the previously best-scoring path is kept (lines 16-20).

\subsection{Shuttling and execution}

Once a set of gates has been assigned to a time step, the algorithm processes all necessary ion movements. Each shuttling step is derived from the previously computed shortest paths between traps: every path is decomposed into individual trap-to-trap transitions, and each transition is stored as a separate shuttle operation. These shuttle operations are represented as nodes in a Directed Acyclic Graph, which encodes the ordering and dependencies among all ion-transport actions.

When a new shuttle is added to the DAG, the algorithm checks all previously recorded shuttles to determine whether dependencies must be introduced. Conflicts arise when two shuttles involve the same ion or traverse a shared junction. In such cases, directed edges are created to enforce the correct execution order. If no conflict exists, the shuttles are considered independent and may be executed in parallel.

During the execution of a time slice, all shuttles required to place qubits in their designated traps are performed by repeatedly extracting the leaf nodes of the shuttle DAG. Each set of leaf shuttles can be executed in parallel, together with the necessary SWAP operations, maximizing movement-level parallelism. Multiple rounds of parallel shuttling may be required until all ions reach their assigned traps.

Once all ion movements for the time slice are completed, all gates scheduled in that slice are executed concurrently. The algorithm then proceeds to the next set of shuttles and gates. This alternating sequence of (i) parallel shuttling rounds followed by (ii) a parallel gate-execution step continues until all operations in the quantum circuit have been completed.
\section{Methodology}
\label{sec:met}

This section describes the methodology followed to evaluate the proposed qubit routing algorithm within ion-trap QCCD architectures. It details the simulation framework used as well as the benchmark circuits set, the performance metrics considered, and the topologies under study. Furthermore, it explains the optimization process applied to tune the routing parameters and analyze their impact on fidelity across different circuit structures and device configurations.

\subsection{Simulation Framework and Baseline Comparison}

The experiments presented in this work were performed using the QCCDSim framework~\cite{martonosi}, a simulation tool for ion-trap QCCD architectures. QCCDSim provides detailed modeling of heating effects, execution times, and fidelities, while supporting the integration of custom compilation strategies, such as the qubit routing algorithm proposed here. It also allows the evaluation of different trapped-ion QCCD topologies and performance metrics.

\begin{figure}[t!]
  \centering

    \includegraphics[width=0.87\linewidth]{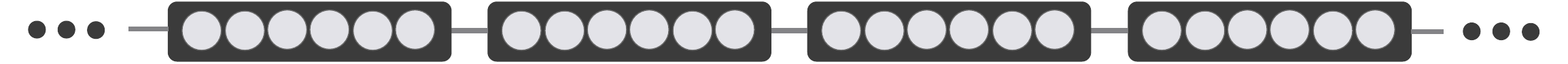}\\[16pt]
    \includegraphics[width=0.7\linewidth]{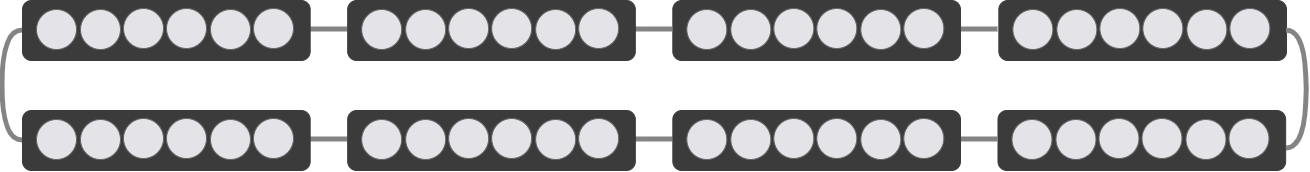}\\[16pt]
\includegraphics[angle=90,width=0.6\linewidth]{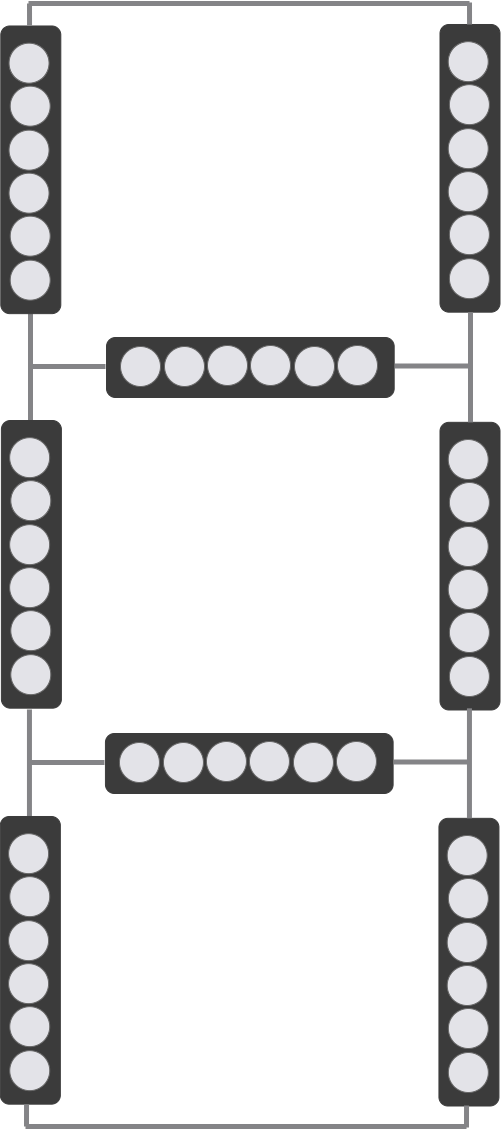}

  \caption{\small{Topologies used in the analysis: linear array (top; L), ring structure (middle; R), and grid topology (bottom; H), each consisting of eight traps with a capacity of six ions.}}

  \label{fig:topology}
\end{figure}

To analyze the proposed qubit routing algorithm, an initial qubit placement and a gate-level Directed Acyclic Graph of each circuit were required. The initial placement was obtained using the one already implemented in the QCCDSim framework, given its adaptability to different topologies, while the DAG representation was generated with Qiskit~\cite{Qiskit}. 

Our qubit routing algorithm was compared against the state-of-the-art approach described in~\cite{muzzle}, which is already implemented within the same framework. This baseline algorithm also assigns scores to traps for qubit placement but does not account for parallelism, making it a suitable point of comparison. Another key difference lies in the execution strategy: the baseline performs operations dynamically (i.e., as soon as one operation finishes, the next begins), whereas our approach organizes execution in distinct “rounds” (i.e. a round of shuttling operations followed by a round of parallel gates execution). This distinction directly affects how the routing scales with circuit structure and topology, motivating a broader performance evaluation.

\subsection{Benchmark Circuits}

To further evaluate and validate the algorithm’s performance, six 40-qubit benchmarks were employed across different QCCD topologies and configurations. The circuit size was fixed to 40 qubits to enable consistent comparison across configurations while supporting the extensive parameter sweep required for weight optimization. This scale is sufficient to expose non-trivial routing congestion and parallelism trade-offs while remaining computationally tractable as analyzed in~\cite{par}. The benchmarks encompass both structured and unstructured quantum circuits, allowing routing performance to be analyzed as a function of circuit structure and highlighting how distinct algorithmic patterns affect qubit routing behavior. The benchmark set includes the Cuccaro Adder (CA), Draper Adder (DA), Quantum Approximate Optimization Algorithm (QAOA), Quantum Fourier Transform (QFT), and two random circuits with the same number of two-qubit gates but different distributions of qubit interactions (RND10 and RND80). 
    \begin{table}[h!]
        \centering
        \vspace{-3pt}

        \vspace{-8pt}
        %{c|ccc|ccc|ccc|c}
        \begin{tabular}{*{5}{c}}
        %\begin{tabular}{*{11}{c}}
            %\cline{2-10}
            \multirow{2}{*}  & \multirow{2}{*}{\textbf{Depth}} & \multirow{2}{*}{\textbf{2q Gates}} & \multirow{2}{*}{\textbf{Av. 2q-Gates/TS}} &\multirow{2}{*}{\textbf{Av. Ion Mov/TS}}\\
            & & & \\ [-1ex]
            \hline
            \textbf{CA}  & 268 & 305 & 1.13  &1.28\\
            \textbf{DA}  & 95 & 590 & 6.21 & 12\\
            \textbf{QAOA}  & 77 & 780 & 10.13 & 19.74\\
            \textbf{QFT}  & 77 & 780 & 10.13 & 19.74\\
            \textbf{RND10}  & 40 & 800 & 20 & 2.95\\
            \textbf{RND80} & 40 & 800 & 20 & 30.1\\
            
            \hline
        \end{tabular}
        \caption{\small{Benchmarks - 40 Virtual Qubits}}
        \label{tab:Benchmarks}
        \vspace{-8pt}
    \end{table}

    Table~\ref{tab:Benchmarks} summarizes key circuit characteristics: the circuit depth (computed using Qiskit~\cite{Qiskit}), the total number of two-qubit gates, the average number of two-qubit gates per time slice (TS, representing the number of operations that can be performed in parallel at a given time in the circuit execution), and the average qubit movement per time slice, which quantifies the ion displacement (shuttling)  in the worst-case scenario where an ion must be moved for each new qubit interaction, as introduced in the work~\cite{par}. It is calculated as follows.

\begin{equation}
    \text{Av.}_{\text{ion\_mov/TS}} = \sum_{i}^N\frac{M_{q_i}}{D},
\end{equation}

\noindent where $N$ is the total number of qubits, $q_i$ represents a qubit from the quantum circuit, $D$ is the circuit depth, and $M_{q_i}$ is the number of movements per qubit (one per interaction with a different qubit).

By observing Table~\ref{tab:Benchmarks}, the structural differences among the benchmark algorithms can be clearly identified. The Cuccaro Adder (CA) exhibits a large circuit depth with a relatively small number of two-qubit gates, resulting in a sequential structure characterized by few two-qubit operations per time slice. Consequently, this algorithm shows the lowest number of qubit interactions per time step, corresponding to minimal ion movement. In contrast, the Draper Adder (DA) presents a considerably smaller depth but increases both the total and average number of two-qubit gates and qubit interactions per time slice, making it suitable for comparison between two distinct structured algorithms. Finally, the Quantum Approximate Optimization Algorithm (QAOA) and Quantum Fourier Transform (QFT) circuits feature a much higher number of two-qubit gates and a lower depth compared with the previously mentioned algorithms, resulting in denser circuits with higher parallelism and greater ion movement per time slice.

In addition, the benchmark set includes two random quantum circuits, designed to test the routing algorithm under scenarios where future operations cannot be predicted. In these random circuits, the number and distribution of qubit interactions can be explicitly controlled. Two configurations were defined: one in which qubit interactions occur mostly between the same pairs of qubits (RND10), and another in which interactions are highly diverse across the algorithm (RND80). These settings lead, respectively, to a very low and a very high average number of ion movements per time slice, thereby allowing the evaluation of routing performance under both localized and highly non-local interaction patterns.

Overall, the benchmark suite is designed to stress different compiler and hardware aspects, including low versus high two-qubit gate density per time slice (CA, DA), regular versus highly interconnected circuit structures (QFT, QAOA), and varying levels of qubit interaction (RND10, RND80).

\subsection{Performance Metrics and QCCD Topologies}

To evaluate the proposed qubit routing algorithm, several metrics were used to characterize the performance. The number of \textit{SWAPs} and \textit{shuttling} operations quantify routing overhead and ion transport cost. The \textit{execution time} accounts for the total duration of all operations, including transport, and gate execution times as modeled in QCCDSim~\cite{martonosi}. Finally, the \textit{fidelity} represents the primary performance indicator, combining the effects of gate errors, ion-chain vibrational motion, and fidelity reduction due to shuttling operations. To incorporate the impact of decoherence, the original QCCDSim fidelity was refined by multiplying it by the qubit coherence factor, as done in previous work~\cite{par}, computed as

\begin{equation}
    C = e^{-\frac{t}{T_2}},
    \label{eqn:coherence}
\end{equation}

\noindent where $t$ denotes the total execution time and $T_2$ represents the qubit coherence time. This correction provides a more realistic estimation of the effective circuit fidelity after execution. Further details on the computation of the base fidelity metric can be found in~\cite{martonosi}. These metrics allow assessing the performance of the qubit routing algorithm, as well as comparing results across different circuit structures and QCCD topologies.

To perform this evaluation under diverse architectural conditions, the execution of the benchmarks was simulated on three distinct topologies: a 1D linear array, a ring structure, and a grid topology (Figure~\ref{fig:topology}). These topologies were selected to represent increasing levels of architectural complexity and trap connectivity. For clarity and generality, the structure of the ion-trap QCCD architecture was simplified, considering only the execution zones while omitting auxiliary regions such as loading or storage zones. Moreover, these configurations are widely adopted in the state of the art: the linear topology has been used in previous works~\cite{martonosi,muzzle}, while the ring structure constitutes a natural extension of the linear one, providing enhanced trap connectivity. This configuration also appears in recent quantum computers such as the Quantinuum H2 QCCD system~\cite{quantinuum2025}, one of the most advanced ion-trap platforms currently available. The grid topology has likewise been investigated in several prior studies~\cite{multiQCCD, efficientcompilationshuttlingtrappedion}, offering higher connectivity and enabling the evaluation of more complex routing scenarios.

\subsection{Parameter Optimization Process}

\begin{figure*}[t] % or [htbp]
  \centering
  \includegraphics[width=\textwidth]{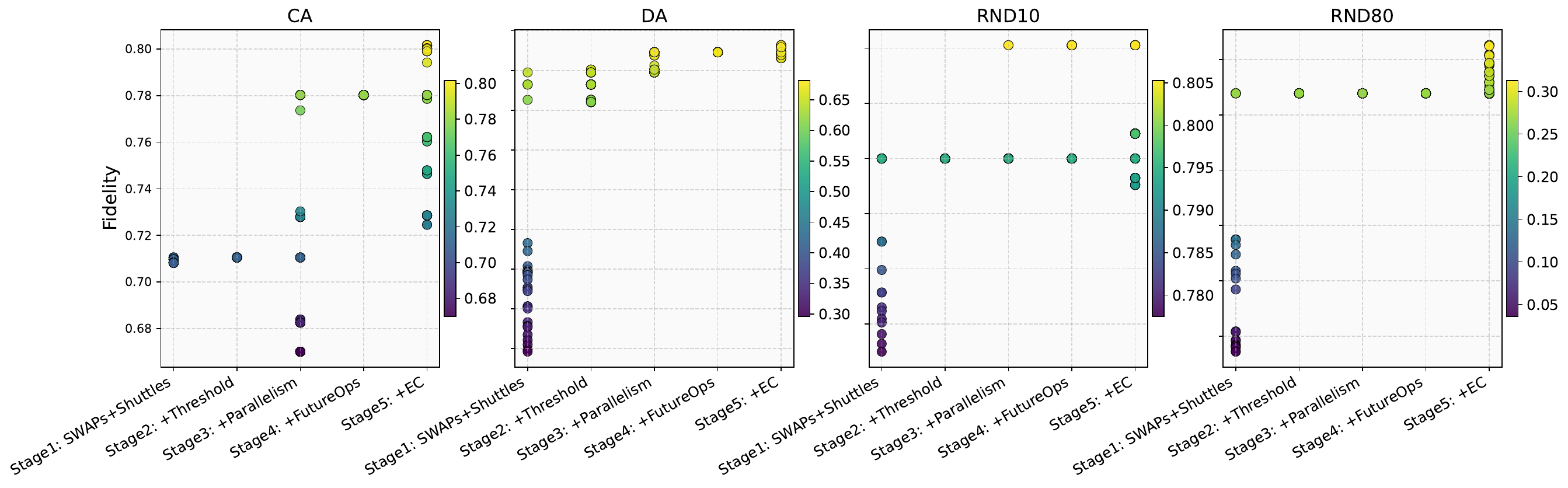} % no .pdf needed
  \caption{\small{Parameter-weight optimization process for different algorithms. First, SWAP and shuttling weights are compared, followed by threshold, operation parallelism, and excess capacity tuning. Each stage optimizes one parameter while fixing the others, progressively improving overall fidelity.}}
  \label{fig:Stages}
\end{figure*}

As previously discussed, the proposed qubit routing algorithm can be configured to adapt to different quantum circuits and device topologies through a set of weighting parameters (Eq.~\ref{ec:TrapScore}). In this work, we systematically explore these parameters, analyze their influence on routing performance, and identify configurations that maximize execution fidelity across different algorithm–architecture scenarios. We decompose this study into five sequential stages aimed at maximizing the fidelity after execution and identifying when, and why, particular settings perform best (Fig.~\ref{fig:Stages}). At each stage, one parameter is swept while the others are held fixed, and the top ten configurations in terms of fidelity are retained and carried forward. It is important to note that the weights in Eq.~\ref{ec:TrapScore} are not normalized since the score components have different numerical scales and physical impact (e.g., accumulated shuttling counts versus trap-capacity indicators); they therefore act as scaling factors to balance their relative influence.

At the beginning of the optimization process, the \textit{SWAP} and \textit{shuttling} parameters are assigned the highest weights, as they have the greatest impact on reducing ion movements, consistent with observations in the state of the art. The threshold parameter is initialized to $-350$, a value consistent with the typical magnitude of trap scores computed from Eq.~\ref{ec:TrapScore}. Since shuttling and SWAP contributions dominate the score and accumulate negative values, this initialization prevents the algorithm from degenerating into purely sequential execution while still allowing sufficiently low-scoring traps to be explored during the first optimization stage.  The exploration proceeds through the following sequential stages, ordered according to the relative impact of the score components: dominant movement-related weights are optimized first, followed by the threshold parameter, whose effective range depends on the resulting score scale, and finally secondary parameters that fine-tune routing behavior.

\begin{itemize}
    \item Stage 1. Compare \textit{SWAP} and \textit{shuttling} weights (approximately 1--65 and 30--180, respectively) with the remaining parameters fixed to one. These two weights span wider ranges because SWAP and shuttling operations dominate resource usage and fidelity loss.
    \item Stage 2. Tune the \textit{threshold} weight (approximately $-350$ to $-60$) using the best SWAP/shuttle settings from Stage~1.  
    \item Stages 3--5. Sequentially optimized \textit{parallelism}, \textit{future operations}, and \textit{excess capacity} (each approximately 1--20), always using the best configuration from the previous stage as the starting point. These parameters require narrower ranges due to their comparatively smaller influence on overall fidelity variation.
\end{itemize}

This staged procedure allows us to assess how each weight influences fidelity across different algorithm types and hardware topologies (including variations in trap capacity and connectivity). When a parameter exhibits no measurable effect in a given scenario, or multiple weights yield indistinguishable fidelities, the value is chosen arbitrarily. Moreover, for the experiments discussed in the next section, the \textit{future-operations} parameter does not meaningfully affect fidelity, and is therefore omitted from the analysis. This limited influence is already visible in Fig.~\ref{fig:Stages}, where varying this weight produces almost no change in the resulting fidelity.

The figure further depicts a representative example of the parameter-weight optimization process. It can be observed that each time a parameter is tuned, the resulting fidelity increases; however, inappropriate weights may also reduce the resulting fidelity. In some cases (for example, parallelism or future-operations weights in this scenario), varying the parameter produces no change in fidelity, indicating that these parameters have little or no impact on the routing process under certain conditions. Overall, fidelity tends to improve progressively as the parameters are refined across the five stages.

\section{Results}
\label{sec:res}

\begin{figure}[t!]
  \centering

  \includegraphics[width=\columnwidth]{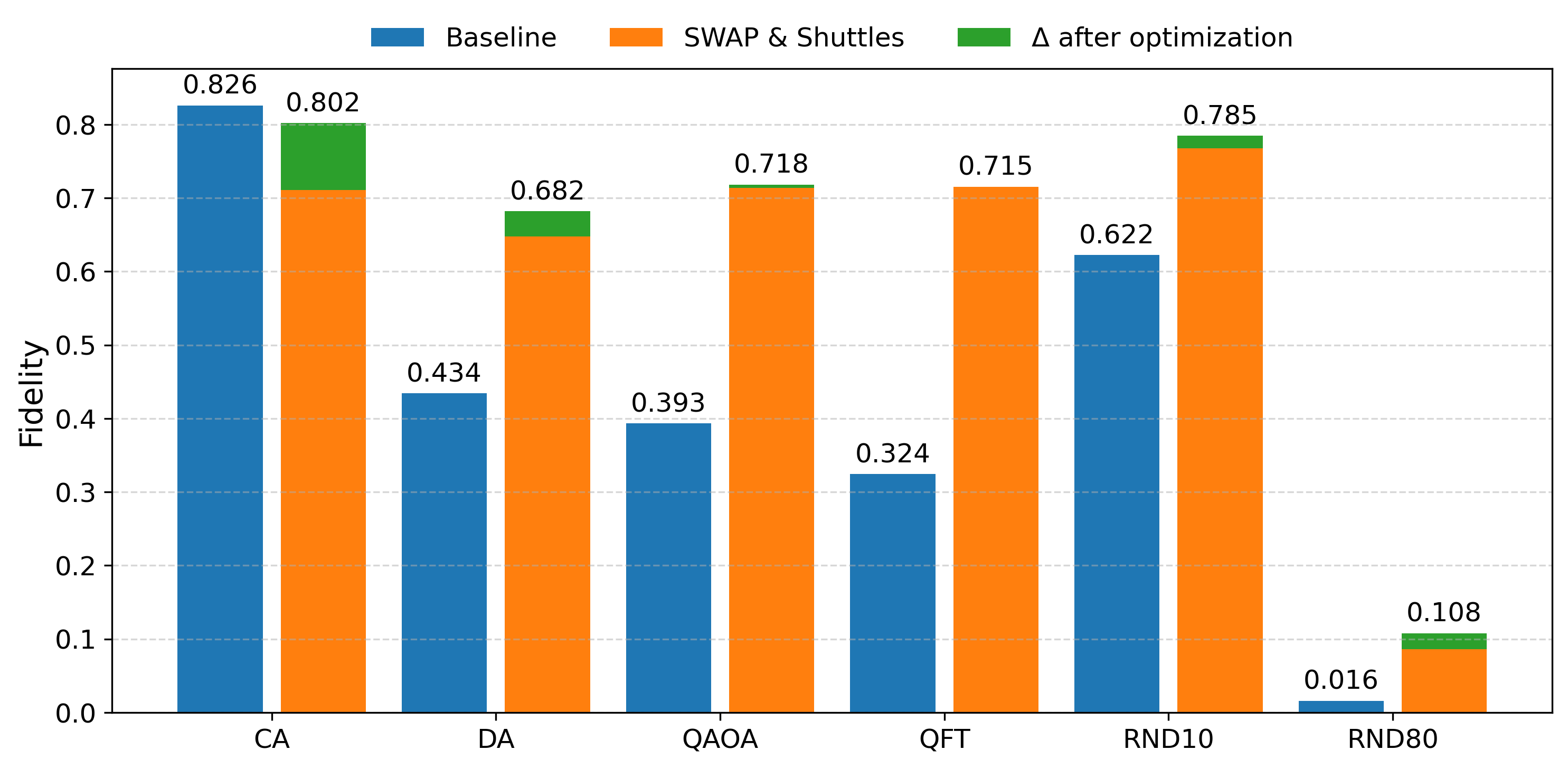}

  \caption{\small{Comparison of proposed qubit routing algorithm against the state of the art (blue). The proposed method first optimizes SWAPs and shuttles (orange) and subsequently tunes the remaining parameters (green).}}
  \label{fig:state_Lin}
\end{figure}
In this section, we demonstrate how the proposed routing algorithm adapts to various quantum circuits, ion distributions across traps, and different trap connectivity topologies. The results are compared against a state-of-the-art qubit routing~\cite{muzzle} technique to evaluate its performance. Additionally, we analyze the influence of the algorithm’s internal weight parameters to identify which components have the greatest impact under different architectural and circuit conditions.

\subsection{Performance Across Quantum Algorithms}
\subsubsection{Comparison with the State of the Art}

\begin{table*}[h]
    \centering
    \vspace{-3pt}

    \resizebox{\textwidth}{!}{
    \vspace{-5pt}
    \begin{tabular}{*{15}{c}}
       \cline{2-11}
        \multirow{-1}{*}{} & \multicolumn{5}{|c|}{\textbf{parallel}} & \multicolumn{5}{c|}{\textbf{baseline}} \\
        \cline{2-11}
         & & & & & & & & & \\ [-1.5ex]
        & \textbf{Shuttles} & \textbf{SWAPs} & \textbf{Ex.Time(ms)} & \textbf{C.Time(s)} & \textbf{Fidelity} & 
        \textbf{Shuttles} & \textbf{SWAPs} & \textbf{Ex.Time(ms)} & \textbf{C.Time(s)} & \textbf{Fidelity} &
        \textbf{$\Delta SH$} & \textbf{$\Delta T$} & \textbf{$\Delta F$} \\
        \hline
        \textbf{RND10} & 1640 & 142 & 46.07 & 207.5 & 0.785 &
                         4600 & 398 & 174.96 & 4.2& 0.622 &
                         \textbf{+64.35} & \textbf{+73.67} & \textbf{+26.08} \\
        \textbf{RND80} & 17930 & 1670 & 441.15 & 397.3& -- &
                         26580 & 2437 & 1007.26 & 57.6 & -- &
                         \textbf{+32.54} & \textbf{+56.20} & -- \\
        \textbf{CA} & 4090 & 294 & 175.32 & 3.64 & 0.802 &
                      2960 & 257 & 154.88 & 0.66 & 0.826 &
                      -38.18 & -13.20 & -2.90 \\
        \textbf{DA} & 4420 & 401 & 105.08 & 19 & 0.682 &
                      9050 & 829 & 363.48 & 7.9 & 0.434 &
                      \textbf{+51.16} & \textbf{+71.09} & \textbf{+57.05} \\
        \textbf{QAOA} & 2530 & 236 & 40.33  & 55 & 0.718 &
                        8610 & 783 & 332.29  &8.36& 0.393 &
                        \textbf{+70.62}& \textbf{+87.86} & \textbf{+82.60} \\
        \textbf{QFT} & 2530 & 239 & 40.53   & 58.5 & 0.715 &
                       9800 & 928 & 387.47   & 13.8 & 0.324 &
                       \textbf{+74.18} & \textbf{+89.54} & \textbf{+120.59} \\
        \hline
    \end{tabular}
    }
    \caption{\small{Performance comparison between the baseline routing method~\cite{muzzle} and the proposed parallelism-aware routing algorithm on a 1D-linear topology (8 traps, 6 ions per trap). The table reports the number of shuttles (SH), SWAPs (SW), simulation execution time (Ex.Time), compilation time (C.Time), and execution fidelity. The $\Delta$ columns indicate the relative improvement of the proposed method with respect to the baseline.}}
    \label{tab:Icomparison}
\end{table*}

\begin{figure*}[t]
  \centering

  \includegraphics[width=0.85\textwidth]{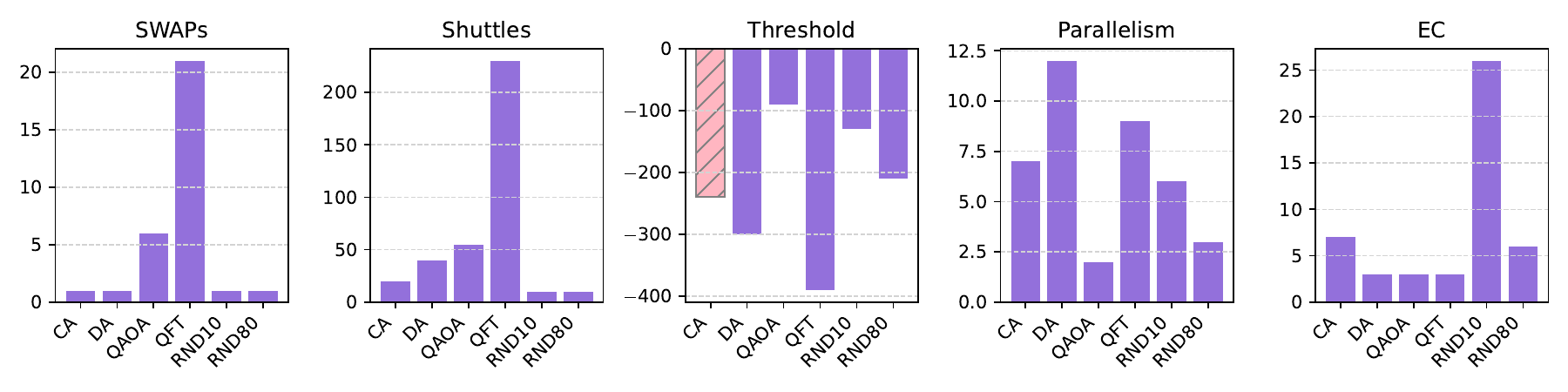}

  \caption{\small{Weight parameters. Pink indicates parameters with no impact on the routing process, assigned randomly.}}
  \label{fig:bench_comp}
\end{figure*}

The qubit routing algorithm is first evaluated using a diverse set of quantum algorithms implemented on a one-dimensional (1D) linear array, as illustrated in Fig.~\ref{fig:topology}. The array consists of eight traps, each capable of holding six ions. This configuration was selected for two main reasons. First, the state-of-the-art approach used for comparison was designed and tested on a 1D linear topology~\cite{muzzle}, allowing for a consistent and fair performance evaluation. Second, a more complex ion distribution was intentionally adopted, characterized by a relatively low number of ions per trap and a larger number of traps. This configuration increases the number of required ion shuttling operations across the device, thereby demanding higher routing precision to minimize total ion movement and enhance overall execution efficiency.

The comparative results between the proposed routing algorithm and the state-of-the-art baseline are presented in Fig.~\ref{fig:state_Lin} and Table~\ref{tab:Icomparison}. Except for the Cuccaro Adder (CA), the proposed method consistently outperforms the baseline, achieving fewer shuttles and SWAPs, shorter simulation execution times, and higher post-execution fidelities. Its compilation time is higher, although this overhead can be substantially reduced with a vectorized implementation of the routing procedure. In Table~\ref{tab:Icomparison}, fidelity results for RND80 are omitted due to being too low to provide meaningful comparison. The improvement is particularly striking for the QFT benchmark, where the algorithm reduces SWAP and shuttling operations by \textit{74\%}, decreases simulation execution time by \textit{89.5\%}, and increases final fidelity by \textit{120.6\%}. These results demonstrate that the proposed routing strategy is effective not only for structured circuits where operation patterns can be anticipated and ion movement optimized, but also for worst-case scenarios such as random circuits.

Figure~\ref{fig:state_Lin} also shows that the SWAP and shuttling weights are the dominant parameters for maximizing fidelity, as previously discussed in the methodology. It also reveals that the influence of the remaining parameters varies across benchmarks, indicating that each algorithm responds differently to weight optimization.These observations motivate a more detailed analysis of how weight tuning interacts with circuit structure, which is examined in the following subsection.

\subsubsection{Influence of Weight Configuration and Algorithm Structure}
Beyond the performance comparison with the state of the art, it is important to understand why different algorithms benefit differently from weight optimization. To this end, we analyze how the optimal weight configurations depend on the structural properties of each circuit.

\begin{figure*}[t]
  \centering

  \includegraphics[width=\textwidth]{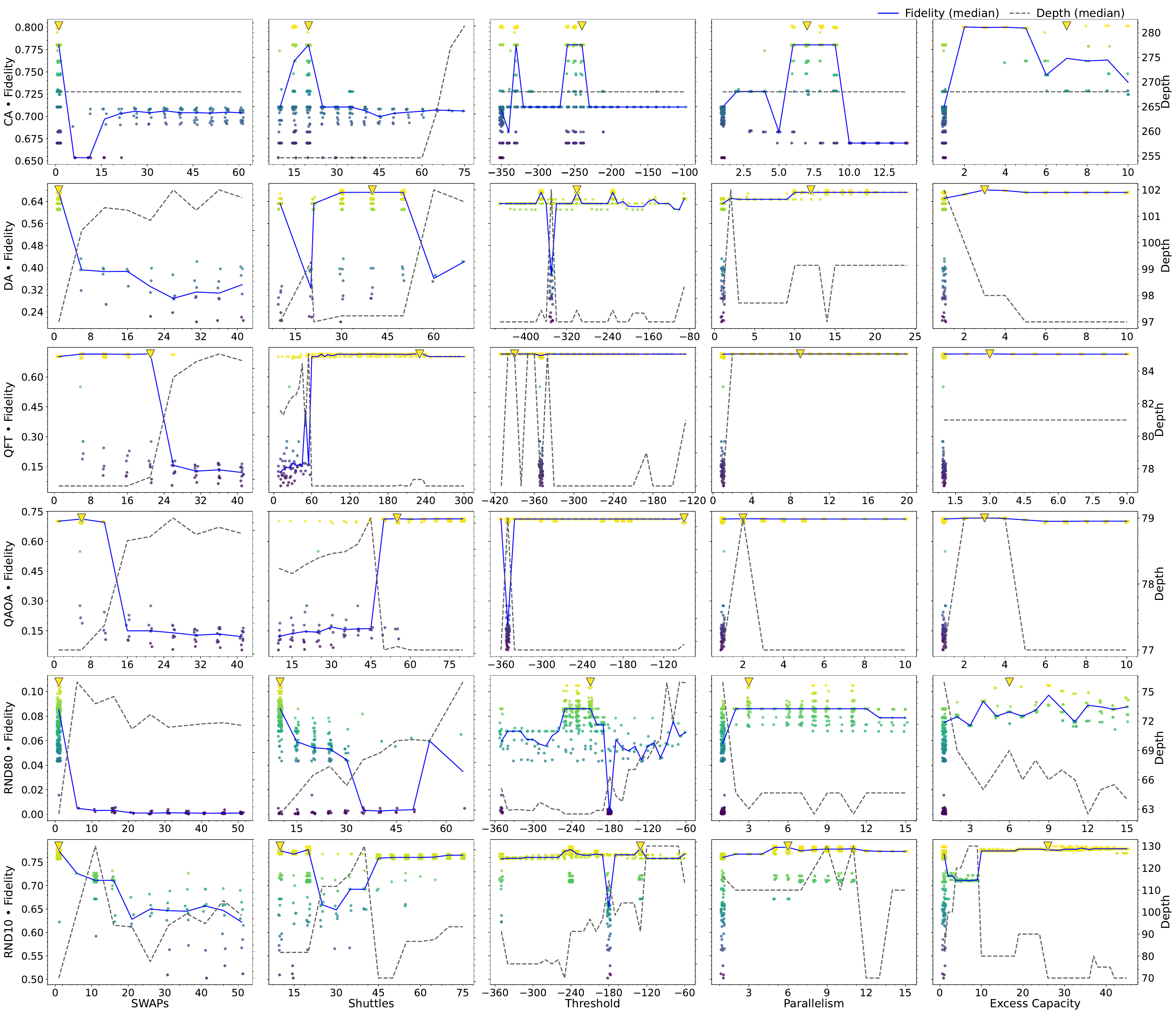}

  \caption{\small{Weight-parameter distribution for all quantum algorithms.
The blue line marks the median parameter weight, the yellow triangle indicates the weight yielding maximum fidelity, and the dashed grey line shows the circuit depth.}}
  \label{fig:big}
\end{figure*}

For instance, the Cuccaro Adder (CA) achieves the largest fidelity improvement after parameter tuning, followed by the Draper Adder (DA). This behavior aligns with their circuit structure: both benchmarks have a low average number of two-qubit gates per time slice, which reduces routing pressure and allows the remaining parameters, particularly EC and Parallelism, to play a more significant role once SWAP and shuttling costs are controlled. This trend is reflected in Fig.~\ref{fig:bench_comp}, where CA and DA assign comparatively higher weights to these parameters relative to the other quantum algorithms.

In contrast, QAOA and QFT experience the smallest improvement from tuning the remaining parameters. These circuits exhibit a high density of two-qubit gates per time slice, resulting in substantial ion-transport activity  across the device. Consequently, the reduction of SWAP and shuttling operations becomes the dominant contributor to fidelity. As Fig.~\ref{fig:bench_comp} shows, the optimal configurations for QAOA and QFT place the highest emphasis on shuttling and SWAP weights, leaving little margin for other parameters to meaningfully influence fidelity.

The behavior of the random circuits (RND10 and RND80) falls between the structured algorithms (CA, DA) and the dense ones (QAOA, QFT). While SWAP and shuttling remain the dominant factors, these circuits also benefit from tuning EC and Parallelism, as reflected in their intermediate weight values in Fig.~\ref{fig:bench_comp}.

With respect to the threshold parameter, the QFT benchmark exhibits the highest value. This is consistent with its large SWAP and shuttling weights: a high threshold prevents the algorithm from accepting low-scoring trap assignments, reducing unnecessary ion movement and avoiding overly parallel execution, which would otherwise harm fidelity. In contrast, QAOA shows the lowest threshold value, reflecting its smaller weights for the remaining parameters and the overall lower scoring range. Interestingly, QFT and QAOA contain the same number of two-qubit gates, yet their optimal thresholds differ substantially. This indicates that routing sensitivity depends not only on gate density per time slice but also on the circuit structure and interaction pattern.

Additionally, it can be observed that the threshold value has no influence for the Cuccaro Adder (CA): within the tested range, all feasible parallelism is already achieved regardless of the threshold weight. This suggests that CA may require a lower threshold range than the one used for the other benchmarks to have meaningful impact. This interpretation is consistent with previous work~\cite{par}, which shows that CA exhibits smaller performance gains when increasing parallelism compared with the other algorithms. A similar trend appears in our fidelity results, which may also help explain CA’s slightly weaker performance relative to the baseline approach, whose execution model is more sequential than the proposed one.

Figure~\ref{fig:big} provides a complementary perspective to Fig.~\ref{fig:bench_comp} by visualizing the full distribution of explored weight configurations and their associated fidelities, rather than only the optimal setting. This representation reveals the sensitivity of each algorithm to parameter tuning. While QAOA and QFT exhibit narrow high-fidelity regions concentrated around strong SWAP and shuttling weights, indicating high sensitivity to movement-related parameters, CA and DA display broader high-fidelity regions, suggesting the existence of multiple competitive configurations. 

The random circuits also exhibit broader high-fidelity regions, indicating that multiple weight configurations can achieve competitive performance. This variability reflects the unstructured nature of their qubit interactions and suggests lower sensitivity to precise parameter tuning compared with highly structured algorithms.

The figure further reveals that the configuration yielding the highest fidelity does not always correspond to the one with minimum execution depth. In several cases, aggressively reducing depth increases shuttling and SWAP operations, ultimately degrading fidelity. This observation highlights that minimizing circuit depth does not always correlate with maximizing fidelity, and that an appropriate balance between execution time and ion-movement cost is necessary.

In summary, through the introduction of parallelism into the routing procedure, the proposed method achieves higher fidelity and lower transport overhead than the baseline qubit routing method. The dependence of the optimal weights on circuit structure highlights the importance of adaptive parameter tuning. The next section extends this analysis by varying the number of traps and ion distributions, allowing us to assess how architectural constraints further influence routing effectiveness.

\subsection{Architectural Sensitivity Analysis}
% in the document
\begin{figure*}[!t]
  \centering

  \includegraphics[width=.95\textwidth]{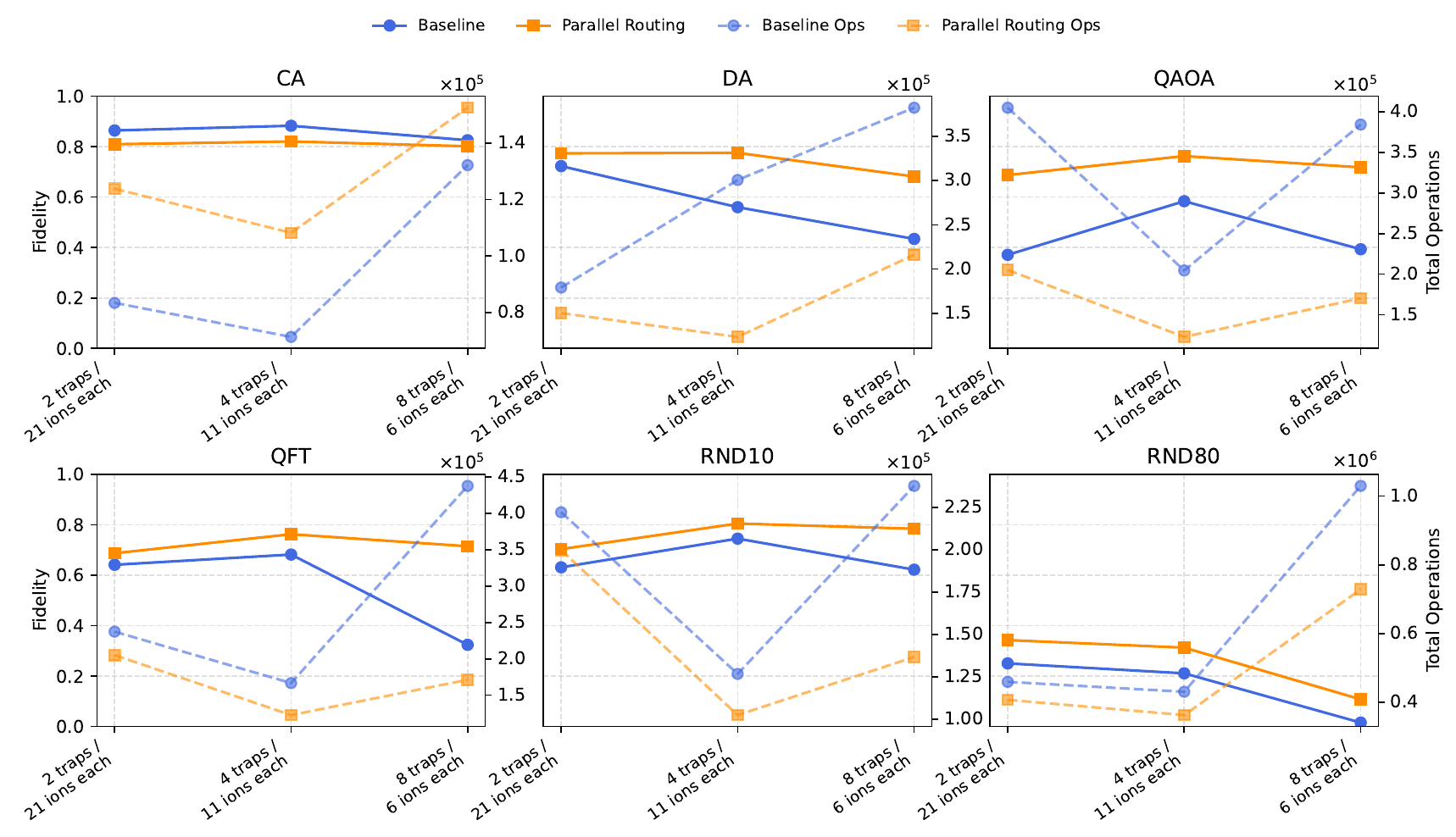}

  \caption{\small{Comparison of the baseline method (blue) and the proposed qubit-routing algorithm (orange), showing fidelity and total operation count (dashed lines) across different trap configurations.}}
  \label{fig:weak_art}
\end{figure*}
\subsubsection{Influence of Trap Capacity and Number of Traps}

After analyzing the behaviour of the benchmark quantum algorithms under the proposed qubit-routing strategy, the next step is to evaluate how the routing performance changes with different trap-topology configurations. To this end, three architectures are evaluated in a 1D linear array topology by varying both the number of traps and their storage capacity, while keeping the total number of qubits constant (40 qubits per algorithm). The evaluated configurations are: (i) 2 traps with 21 ions per trap, (ii) 4 traps with 11 ions per trap, and (iii) 8 traps with 6 ions per trap.

These setups allow us to observe how the routing algorithm responds to architectures with varying degrees of modularity and routing complexity. The configuration with 2 traps and 21 ions per trap represents a routing-simple scenario with limited inter-trap communication, whereas the configuration with 8 traps and 6 ions per trap creates a more modular and routing-intensive environment due to increased trap-to-trap interactions. This comparison highlights how topology complexity impacts routing efficiency and ion-movement costs.

The higher the number of traps and the lower their capacity, the more SWAP and shuttle operations are required to execute all two-qubit gates, as shown in Figure~\ref{fig:weak_art}. It is also noticeable that simpler topologies (e.g., 2 traps with 21 ions each) exhibit only limited fidelity improvement when compared to the state of the art. This is expected, since reduced ion movement provides little opportunity for further optimisation. In contrast, as the topology becomes more complex and the routing procedure more demanding, the advantages of the proposed routing strategy become more pronounced. This trend is reflected in the total number of operations: for the most complex configuration (8 traps with 6 ions per trap), the proposed method reduces total operations by approximately 36\% for DA, 57\% for QAOA, and over 60\% for QFT compared to the baseline (Fig.~\ref{fig:weak_art}). In the highly interactive RND80 benchmark, the reduction remains significant at approximately 25\%. These quantitative differences explain the larger fidelity gains observed in more modular architectures. These results indicate that the relative advantage of the proposed routing strategy increases with architectural modularity and routing complexity, suggesting that its benefits are likely to become more pronounced in larger or more irregular QCCD architectures.

Furthermore, the configuration with 4 traps and a capacity of 11 ions frequently achieves the best overall results. This layout provides an effective balance between parallelism and ion movement, a behaviour also observed in~\cite{par}. Finally, it is noteworthy that only for the random quantum algorithm with high interaction density (RND80) the configuration with 2 traps and 21 ions performs best. In this case, the large number of qubit interactions favours a mostly sequential execution where all ions reside in the same trap, despite the vibrational instabilities introduced by long ion chains, which tend to reduce fidelity.

\begin{figure*}[!t]
  \centering

  \includegraphics[width=\textwidth]{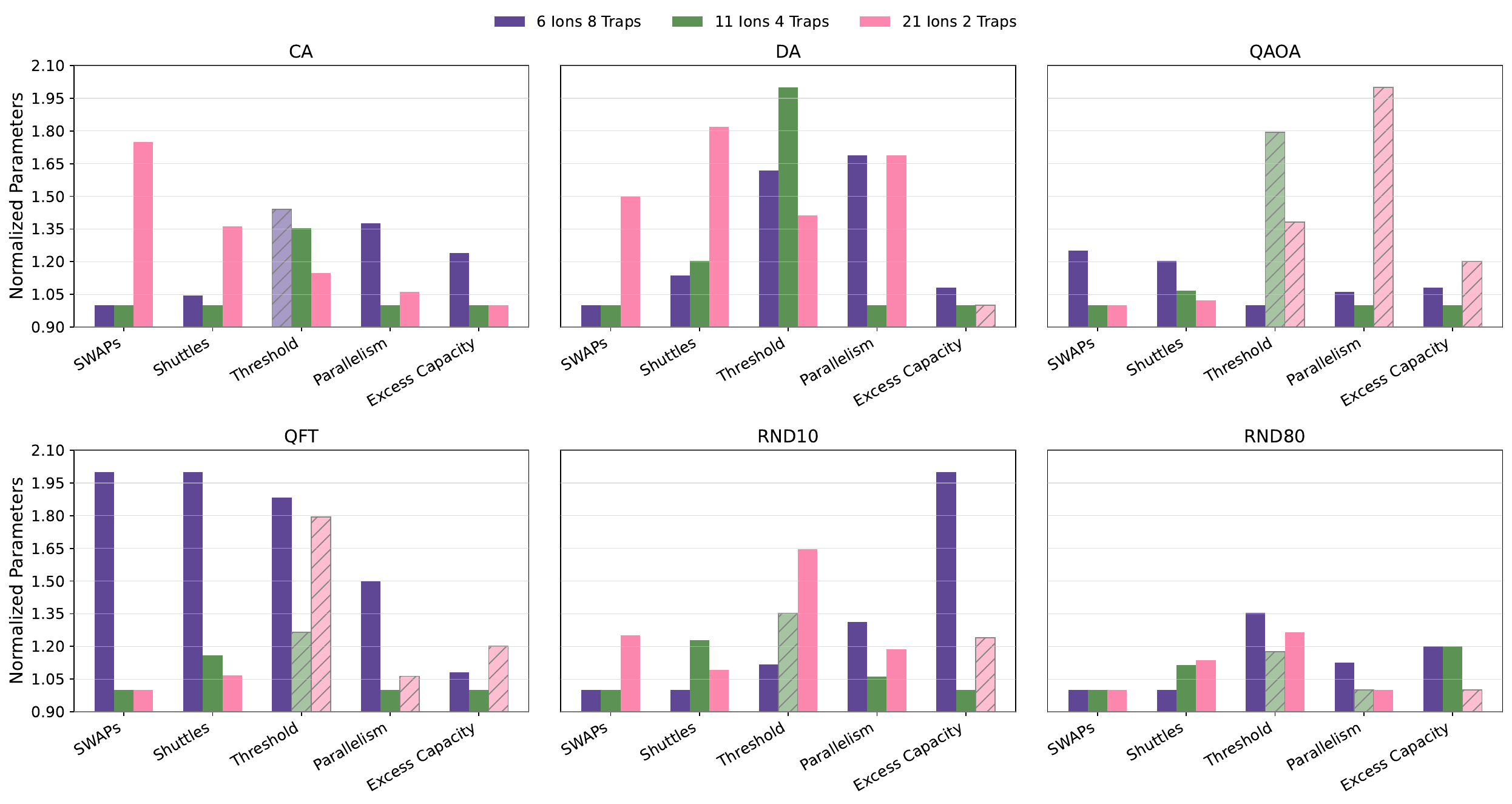}

  \caption{\small{Weight-parameter behavior for a 1D linear-array topology under three ion–trap configurations: 8 traps with 6 ions each (purple), 4 traps with 11 ions each (green), and 2 traps with 21 ions each (pink). Gray lines in bars indicate parameters that had no impact and were assigned randomly.}}
  \label{fig:weak_his}
\end{figure*}

This reduced routing complexity, caused by the simpler topology, is also reflected in the impact of the weight parameters. Figure~\ref{fig:weak_his} shows the normalized parameter weights used during routing, where higher values indicate greater influence on the routing decision. As the routing task becomes easier, the influence of most parameters diminishes, and several of them exhibit no measurable impact on routing performance (grey lines). 

Overall, for the topology with 2 traps and 21 ions per trap, the dominant parameters are the shuttle and SWAP costs. With only two traps available, routing mainly reduces to moving ions within large chains, making the minimization of SWAPs, and the associated shuttles, the primary objective. Parameters such as excess capacity, parallelism, or future operations offer little influence in this setup because the topology provides almost no alternative routing choices. For instance, in the QAOA and QFT cases using the 2-trap, 21-ion configuration, only the SWAP and shuttle weights have a noticeable effect. This behaviour is expected: both algorithms involve many two-qubit gates, making the reduction of ion-movement operations the dominant optimisation target.

Moreover, despite changing the topology and its complexity, each algorithm continues to exhibit distinct parameter-weight profiles. This confirms that the routing behaviour depends not only on the hardware structure but also on the intrinsic interaction patterns of the quantum algorithm.

The results indicate that increasing architectural complexity enhances the relative benefit of the proposed routing approach, yielding larger fidelity gains with respect to the baseline. The analysis so far has considered several quantum algorithms, different numbers of traps, and multiple trap-capacity configurations in linear topologies. The next set of experiments investigates the impact of alternative trap-connectivity patterns on routing behavior.

\subsubsection{Impact of Trap Connectivity}
The last experiment analyzed how the proposed routing algorithm performs under varying trap-connectivity schemes (Fig.~\ref{fig:topology}). This evaluation does not include a comparison against the baseline, since the state-of-the-art method is optimized exclusively for linear architectures, as reported in~\cite{im}. A system of 8 traps with a capacity of 6 ions each is used, and the connectivity between traps is progressively increased. The first configuration corresponds to the 1D linear topology used throughout the previous analysis. Two additional shuttle links are then introduced to form a ring topology. Finally, a grid topology is tested, offering significantly higher connectivity but also requiring more complex routing, as some ion transfers involve traversing two shuttles to reach the destination trap.

\begin{figure}[t!]
  \centering

  \includegraphics[width=\columnwidth]{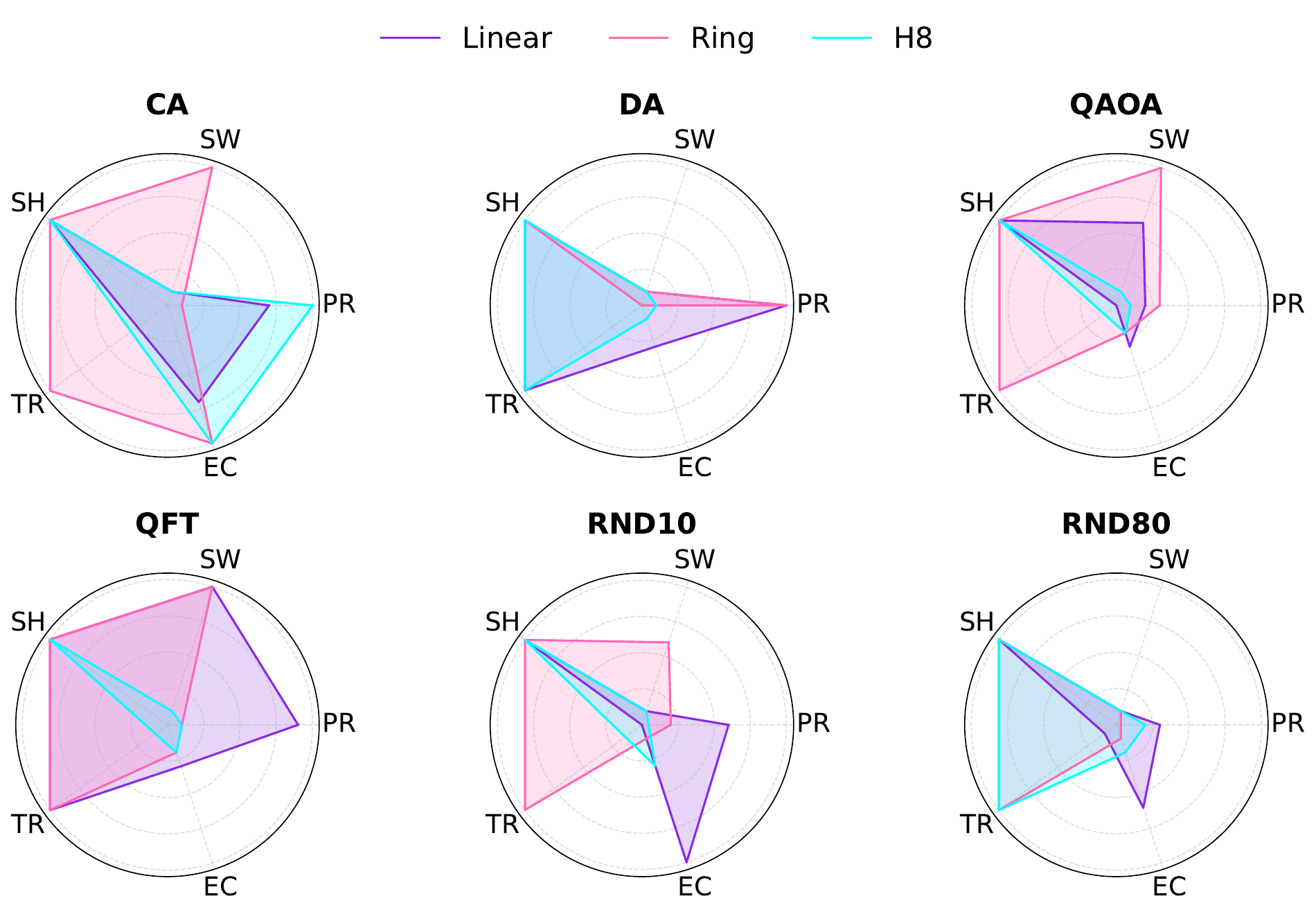}

  \caption{\small{Weight parameter behavior for a 1D linear-array (purple), ring (pink) and grid topology (blue).}}
  \label{fig:topo-sp}
\end{figure}

This change in trap connectivity alters the relative importance of the scoring parameters, as shown in Fig.~\ref{fig:topo-sp}. In the grid topology (H8), the shuttling parameter becomes dominant across all benchmarks, while the remaining parameters have almost no influence on the routing process. This behavior stems from the inherent shuttling complexity of the grid architecture, which forces the routing algorithm to prioritize minimizing ion movement above all else. This effect is particularly evident for QFT and QAOA, where shuttling is effectively the only parameter with significant weight. In these circuits, the combination of dense two-qubit interactions and complex shuttle paths leads to a reduction in fidelity (Table~\ref{tab:topo}), pushing the algorithm to assign maximum weight to shuttling minimization.
For the remaining benchmarks, however, the higher connectivity of the grid topology enables noticeable reductions in both SWAPs and shuttles, resulting in improved post-execution fidelity. Since these algorithms exhibit lower gate density per time slice, they benefit from the additional routing flexibility despite the increased operational complexity of the topology.

When examining the ring topology, the influence of the scoring parameters becomes more balanced. Unlike the grid case, where shuttling dominates, the ring topology assigns significant weight not only to shuttling but also to SWAPs, with the threshold emerging as the next most relevant factor. This reflects the algorithm's tendency to reduce overall ion movement by simultaneously minimizing transport operations and restricting excessive parallelism through a higher minimum score requirement.
In terms of performance, the ring connectivity yields lower fidelity than the grid topology in most benchmarks, yet it consistently outperforms the linear configuration (with the exception of QAOA and QFT, as previously discussed). This shows that the additional connectivity provided by the ring can be exploited by the routing algorithm, even if its benefits are more limited than in the grid architecture.

\begin{table*}[ht]
\centering

\resizebox{\textwidth}{!}{%
\begin{tabular}{lccc|ccc|ccc|ccc|ccc|ccc}
\cline{2-19}

 & \multicolumn{3}{c|}{\cellcolor{lightgray}\textbf{CA}}
 & \multicolumn{3}{c|}{\cellcolor{lightgray}\textbf{DA}}
 & \multicolumn{3}{c|}{\cellcolor{lightgray}\textbf{QAOA}}
 & \multicolumn{3}{c|}{\cellcolor{lightgray}\textbf{QFT}}
 & \multicolumn{3}{c|}{\cellcolor{lightgray}\textbf{RND10}}
 & \multicolumn{3}{c|}{\cellcolor{lightgray}\textbf{RND80}} \\
\cline{2-19}
 & SW & SH & Fid & SW & SH & Fid & SW & SH & Fid & SW & SH & Fid & SW & SH & Fid & SW & SH & Fid \\
\hline
\cellcolor{lightgray}L & 294 & 409 & 0.802 & 401 & 442 & 0.682 & 236 & 253 & 0.720 & 239 & 253 & 0.715 & 142 & 164 & 0.785 & 1670 & 1793 & 0.108 \\
\cellcolor{lightgray}R & 170 & 264 & 0.866 & 306 & 340 & 0.730 & 244 & 269 & 0.720 & 183 & 237 & 0.725 & 115 & 132 & 0.797 & 1180 & 1300 & 0.260 \\
\cellcolor{lightgray}H &  67 & 235 & 0.911 & 293 & 431 & 0.742 & 284 & 393 & 0.689 & 285 & 393 & 0.688 &  74 & 129 & 0.815 &  749 & 1167 & 0.476 \\
\hline

\cellcolor{lightgray}\textbf{$\Delta$} 
 & \textbf{339\%} & \textbf{74\%} & \textbf{12\%} (H)
 & \textbf{37\%} & \textbf{30\%} & \textbf{8\%} (H)
 & \textbf{0\%} & \textbf{0\%} & \textbf{0\%} --
 & \textbf{31\%} & \textbf{7\%} & \textbf{1\%}  (R)
 & \textbf{92\%} & \textbf{27\%} & \textbf{4\%}  (H)
 & \textbf{123\%} & \textbf{54\%} & \textbf{77\%}  (H)\\
\hline
\end{tabular}
} % end resizebox
\caption{\small{Comparison of different trap connectivity configurations. L denotes a linear topology, R a ring, and H a grid. $\Delta$ indicates the improvement relative to the linear topology.}}
\label{tab:topo}
\end{table*}

Lastly, the linear topology achieves the lowest fidelity after execution due to its limited trap connectivity. However, this simplicity also reduces the routing constraints: the weight parameters are not forced to concentrate exclusively on minimizing ion movement, as happens in more complex topologies. As a result, the linear architecture shows greater variability in the optimal parameter values across different quantum algorithms, indicating that the routing strategy can adapt more flexibly to algorithmic structure. In contrast, when the topology becomes more complex, such as in the grid configuration, the routing process must prioritize coping with architectural constraints, and the parameter weights become dominated by movement rather than the characteristics of the quantum algorithm.  Overall, the underlying trap connectivity emerges as a dominant factor governing the routing dynamics and parameter sensitivities of the algorithm.

\section{Conclusions}
\label{sec:con}

This work presented a qubit routing strategy that explicitly incorporates operation-level parallelism in QCCD ion-trap architectures. Through a structured parameter optimization process, the method adapts to both the quantum circuit structure and the underlying device topology. The routing framework jointly considers ion-movement cost, trap capacity constraints, and the balance between sequential and parallel execution, dynamically adjusting decisions according to qubit distribution and trap connectivity. Experimental results show consistent improvements in execution fidelity compared with state-of-the-art routing techniques, with the largest gains observed in highly modular and routing-intensive architectures, achieving an average improvement of 56\% and up to 120\% in the best cases.

Overall, the results highlight the importance of considering both architectural constraints and parallel execution when designing routing strategies for modular trapped-ion quantum processors. Future work will focus on automating the selection of routing parameters to avoid exhaustive sweeps and on extending the framework to support parallel gate operations within the same trap while accounting for crosstalk effects.2
\section*{Acknowledgments}

The authors acknowledge financial support from the European Union’s Horizon Europe research and innovation program through the project Quantum Internet Alliance under grant agreement No. 101102140.

\bibliographystyle{ieeetr}
\bibliography{bib/main}
%\printbibliography

\end{document}